\RequirePackage[2020-02-02]{latexrelease}
\documentclass[pre,twocolumn]{revtex4}
\usepackage{epsfig}
\usepackage{bm}
\usepackage{amsmath}

\newcommand{\bea}{\begin{eqnarray}}
\newcommand{\eea}{\end{eqnarray}}

\begin{document}

\title{Small-scale magnetohydrodynamic dynamos: from deterministic chaos to turbulence}

\author{A. Bershadskii}

\affiliation{
ICAR, P.O. Box 31155, Jerusalem 91000, Israel
}

\begin{abstract}

  It is shown, using results of numerical simulations, and geophysical and solar observations, that the transition from deterministic chaos to hard turbulence in the magnetic field generated by the small-scale MHD dynamos occurs through a randomization process. This randomization process has been described using the notion of distributed chaos and the main parameter of distributed chaos has been used for quantifying the degree of randomization. The dissipative (Loitsianskii and Birkhoff-Saffman integrals) and ideal (magnetic helicity) magnetohydrodynamic invariants control the randomization process and determine the degree of randomization in different MHD flows, directly or through the Kolmogorov-Iroshnikov phenomenology (the magneto-inertial range of scales as a precursor of hard turbulence). Despite the considerable differences in the scales and physical parameters, the results of numerical simulations are in quantitative agreement with the geophysical and solar observations in the frames of this approach. The Hall magnetohydrodynamic dynamo has been also briefly discussed in this context.
  
\end{abstract}

\maketitle

\section{Introduction}  
  
    The concept of smoothness can be functional for the quantitative classification of the non-laminar regimes in magnetohydrodynamics (and in fluid dynamics in general) according to their randomness. The spectral analyses can be used for this purpose. Namely, the  stretched exponential spectra are typical for the smooth magnetohydrodynamics
\begin{equation}
E(k) \propto \exp-(k/k_{\beta})^{\beta}.  
\end{equation}
 here $1 \geq \beta > 0$ and $k$ is the wavenumer. The value $\beta =1$ 
\begin{equation}
E(k) \propto \exp(-k/k_c),  
\end{equation} 
is typical for deterministic chaos (see Refs. \cite{fm}-\cite{kds}). \\

  For $1 > \beta$ the dynamics is still smooth but not deterministic (and will be called the distributed chaos, see below for clarification of the term).  It can be also considered as a soft turbulence \cite{wu}.
 
  The non-smooth (hard turbulence \cite{wu}) dynamics is typically characterized by the power-law (scaling) spectra. \\
  
  In this approach, the value of the $\beta$ can be considered as a proper measure of randomization. Namely, the further the value of the parameter $\beta$ is from the deterministic $\beta =1$ (i.e. smaller the $\beta$) the randomization is stronger. \\

 Let us consider two examples. The first example can be taken from a direct numerical simulation (DNS) reported in a paper Ref. \cite{sch}. Figure 1 shows the magnetic energy spectra generated by a small-scale incompressible saturated MHD dynamo at the magnetic Prandtl number $Pr_m =10$ and different values of the Reynolds number (shown in the figure). The spectral data were taken from Figure 33a of the Ref. \cite{sch} (see below for a more detailed description of this DNS). 
 
 The dashed curves in Fig. 1 indicate the best fits corresponding to the stretched exponentials Eq. (1) and the dotted arrow indicates the position of the characteristic wavenumber $k_c$ from Eq. (2).\\
 
\begin{figure} \vspace{-0.2cm}\centering
\epsfig{width=.46\textwidth,file=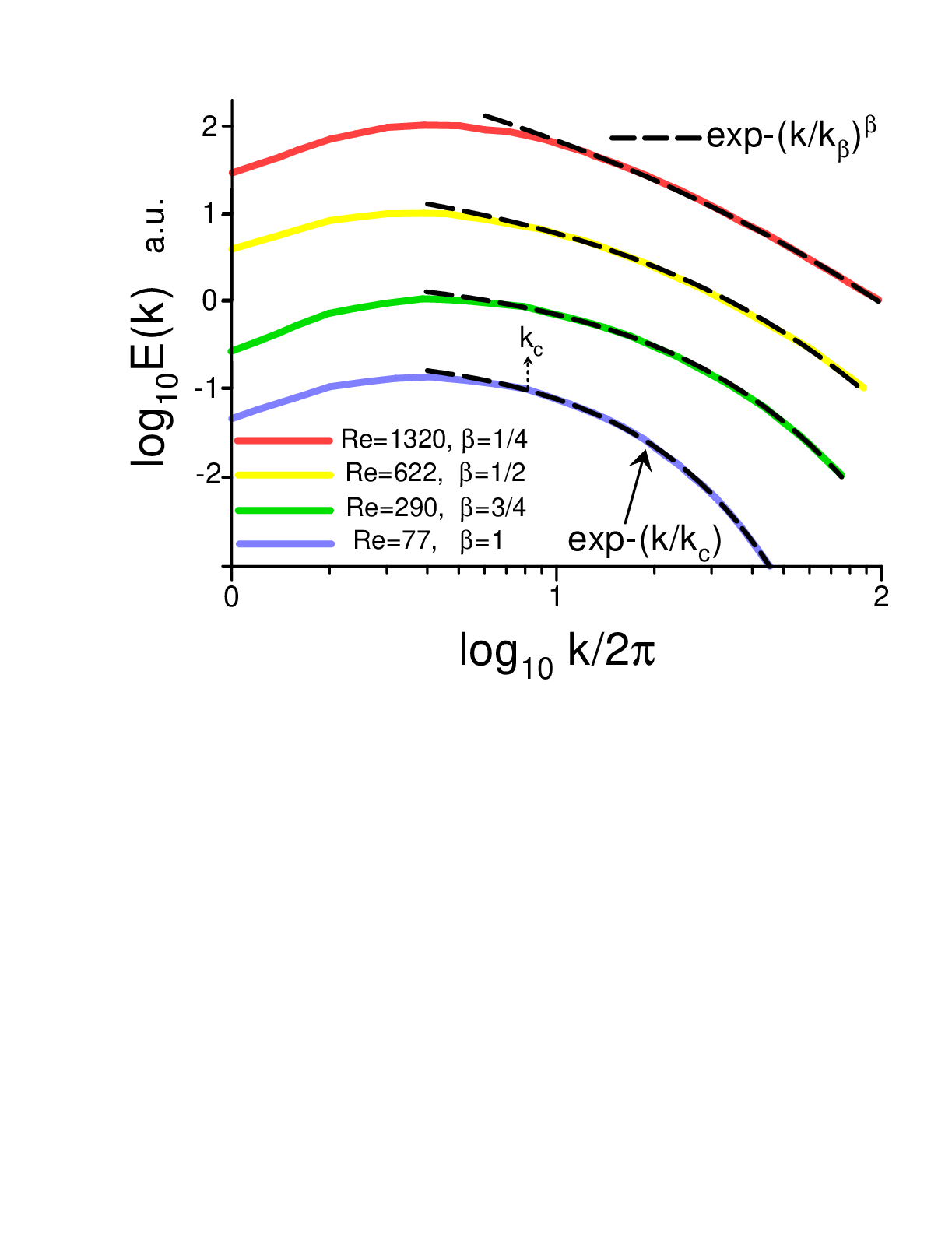} \vspace{-5cm}
\caption{Magnetic energy spectra generated by a small-scale incompressible saturated MHD dynamo at different values of the Reynolds number and at magnetic Prandtl number $Pr_m =10$ (direct numerical simulations). The spectra are vertically shifted for clarity.} 
\end{figure}

 The values of the parameter $\beta$ are decreasing when the value of the Reynolds number is increasing (starting from $\beta = 1$, i.e. from the deterministic chaos at a low Reynolds number) indicating an increasing randomization with the Reynolds number. This trend conforms to physical expectations.\\
 
  The second example can be taken from an observation of the evolution of the magnetic field in a rather large emerging solar active region AR NOAA 11726. The measurements were provided by Helioseismic and Magnetic Imager located on board of the Solar Dynamics Observatory (SDO/HMI).  Figure 2 shows the magnetic power spectra for the emerging active region AR NOAA 11726 at different times of its evolution. The spectral data were taken from Fig. 2a of a recent paper Ref. \cite{kka}. The time moment $t_0$ corresponds to the start moment of the emerging process and $t_0 < t_1 < t_2 < t_3 $ up to the beginning of the decline stage. The dashed curves in Fig. 2 indicate the best fits corresponding to the stretched exponentials Eq. (1) and the dotted arrow indicates the position of the characteristic wavenumber $k_c$ for the deterministic chaos (as in Fig. 1).\\

  One can see that the randomness of the magnetic field increases as the emergence of the active region proceeds (starting from $\beta = 1$ at $t_0$, i.e. from the deterministic chaos at the beginning of the emergence process). This trend also conforms to physical expectations.\\
   
   In the next sections, the observed values of $\beta$ will be related to the magnetohydrodynamic invariants. These relationships allow a deeper understanding of the physics of randomization at the chaotic/turbulent dynamo processes.

\begin{figure} \vspace{-0.65cm}\centering
\epsfig{width=.46\textwidth,file=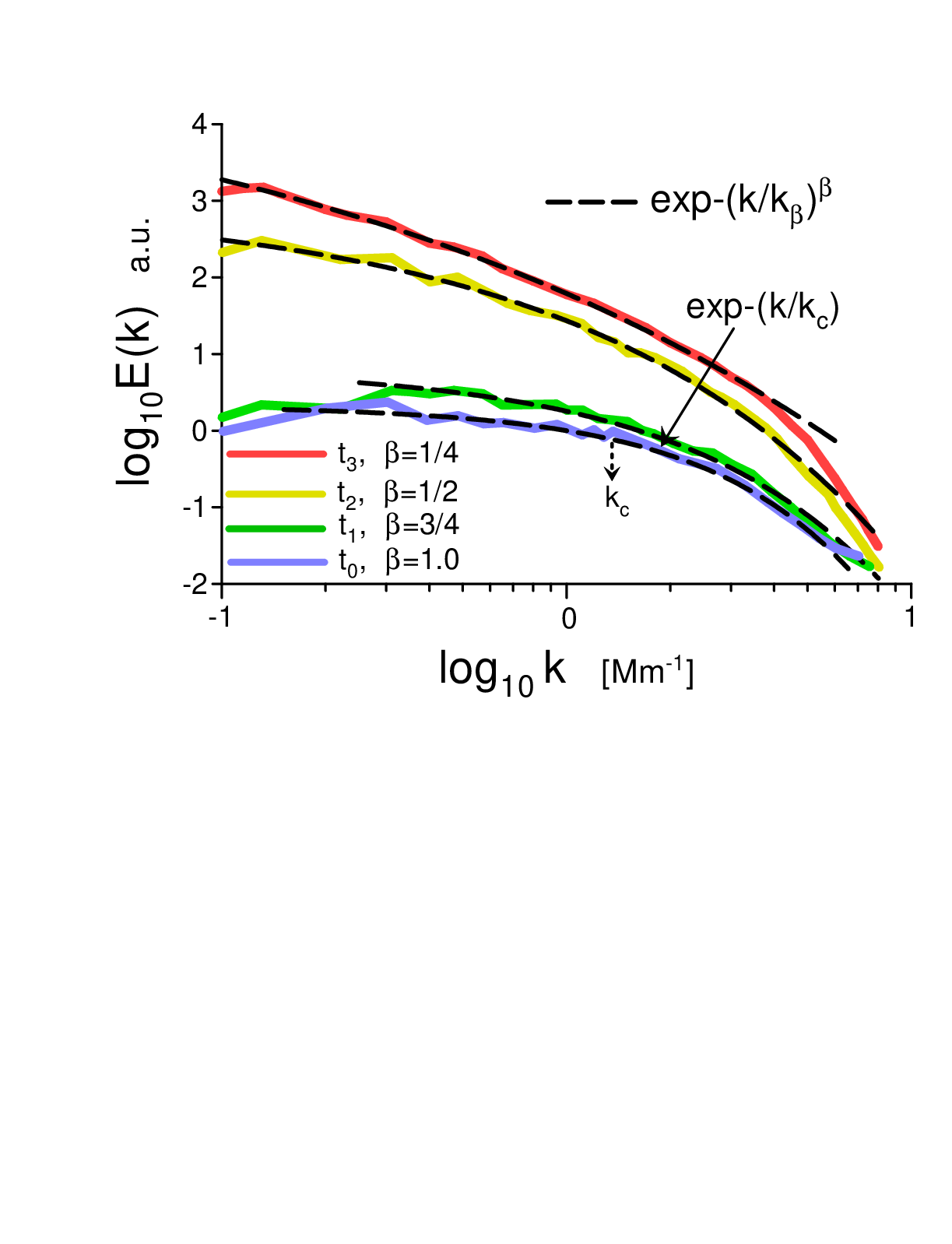} \vspace{-4.8cm}
\caption{ Magnetic energy spectra for an emerging solar active region AR NOAA 11726 at different times of its evolution.} 
\end{figure}
   
\section{Magnetic helicity and distributed chaos}  
 
 \subsection{Magnetic helicity}   
 
 The ideal magnetohydrodynamics has three quadratic (fundamental) invariants: total energy, cross and magnetic helicity \cite{mt}. \\
  
  The magnetic helicity is
\begin{equation}
 h_m = \langle {\bf a} {\bf b} \rangle  
\end{equation}
where the fluctuating magnetic field is ${\bf b} = [{\nabla \times \bf a}]$ ($\nabla \cdot {\bf a} =0$), and a spatial average is denoted as $\langle ... \rangle$. Both fluctuating ${\bf a}$ and ${\bf b}$ have zero means, and $\nabla \cdot {\bf a} =0$. \\

 The presence of a uniform magnetic field ${\bf B_0}$ violates the invariance of magnetic helicity. A modified magnetic helicity was introduced in the paper Ref. \cite{wg} in the form
\begin{equation}
 \hat{h} = h_m + 2{\bf B_0}\cdot \langle {\bf A}  \rangle 
 \end{equation}
here ${\bf B} = {\bf B_0} + {\bf b}$, ${\bf A} = {\bf A_0} +{\bf a}$, and ${\bf b} = [{\nabla \times \bf a}]$. It can be shown that at rather weak restrictions on the boundary conditions in the ideal magnetohydrodynamics \cite{wg} (see also Ref. \cite{shebalin})
\begin{equation}
 \frac{d \hat{h}}{d t} =  0, 
 \end{equation}
i.e. the modified magnetic helicity  $\hat{h} $ is an ideal magnetohydrodynamic invariant in the presence of a uniform magnetic field.\\

\subsection{Distributed chaos in magnetic field driven by magnetic helicity} 
   
     A change of deterministically chaotic system parameters can result in the random fluctuations of the characteristic scale $k_c$ in Eq. (2). In this case, one has to use an ensemble averaging in order to compute the magnetic power spectra:
\begin{equation}
E(k) \propto \int_0^{\infty} P(k_c) \exp -(k/k_c)dk_c 
\end{equation}    
with a probability distribution $P(k_c)$ characterizing the random fluctuations of the characteristic scale $k_c$. Therefore, the corresponding smooth non-deterministic chaotic dynamics will be called `distributed chaos'.\\

    It is well-known that in a weakly non-ideal case magnetic helicity is still almost conserved (since there are no processes that can effectively transfer it to the resistive scales) while the magnetic and total energy are efficiently dissipated, i.e. magnetic helicity is an adiabatic invariant in weakly non-ideal MHD. Therefore one can find the probability distribution $P(k_c)$ for the fluid (or plasma) dynamics dominated by the magnetic helicity using the dimensional considerations and a scaling relationship 
\begin{equation}
B_c \propto |h_m|^{1/2} k_c^{1/2}   
\end{equation}
relating the characteristic value of the magnetic field $B_c$ to the characteristic value of the wavenumber $k_c$ through magnetic helicty $|h_m|$ (as an adiabatic invariant).\\

   If  the positive variable $B_c$ has the half-normal probability distribution $P(B_c) \propto \exp- (B_c^2/2\sigma^2)$ \citep{my} (i.e. a normal distribution with zero mean truncated to only have nonzero probability density for positive values of its argument: if $B$ is a normal random variable,  then $B_c = |B|$ has a half-normal distribution \cite{jkb}), then it follows from Eq. (7) that the characteristic value of the wavenumber $k_c$ has the chi-squared ($\chi^{2}$) distribution
\begin{equation}
P(k_c) \propto k_c^{-1/2} \exp-(k_c/4k_{\beta})  
\end{equation}
here $k_{\beta}$ is a new constant. 

   Substituting Eq. (8) into Eq. (6) we obtain
\begin{equation}
E(k) \propto \exp-(k/k_{\beta})^{1/2}  
\end{equation}

\subsection{Spontaneous breaking of local reflectional symmetry}

\begin{figure} \vspace{-0.6cm}\centering 
\epsfig{width=.45\textwidth,file=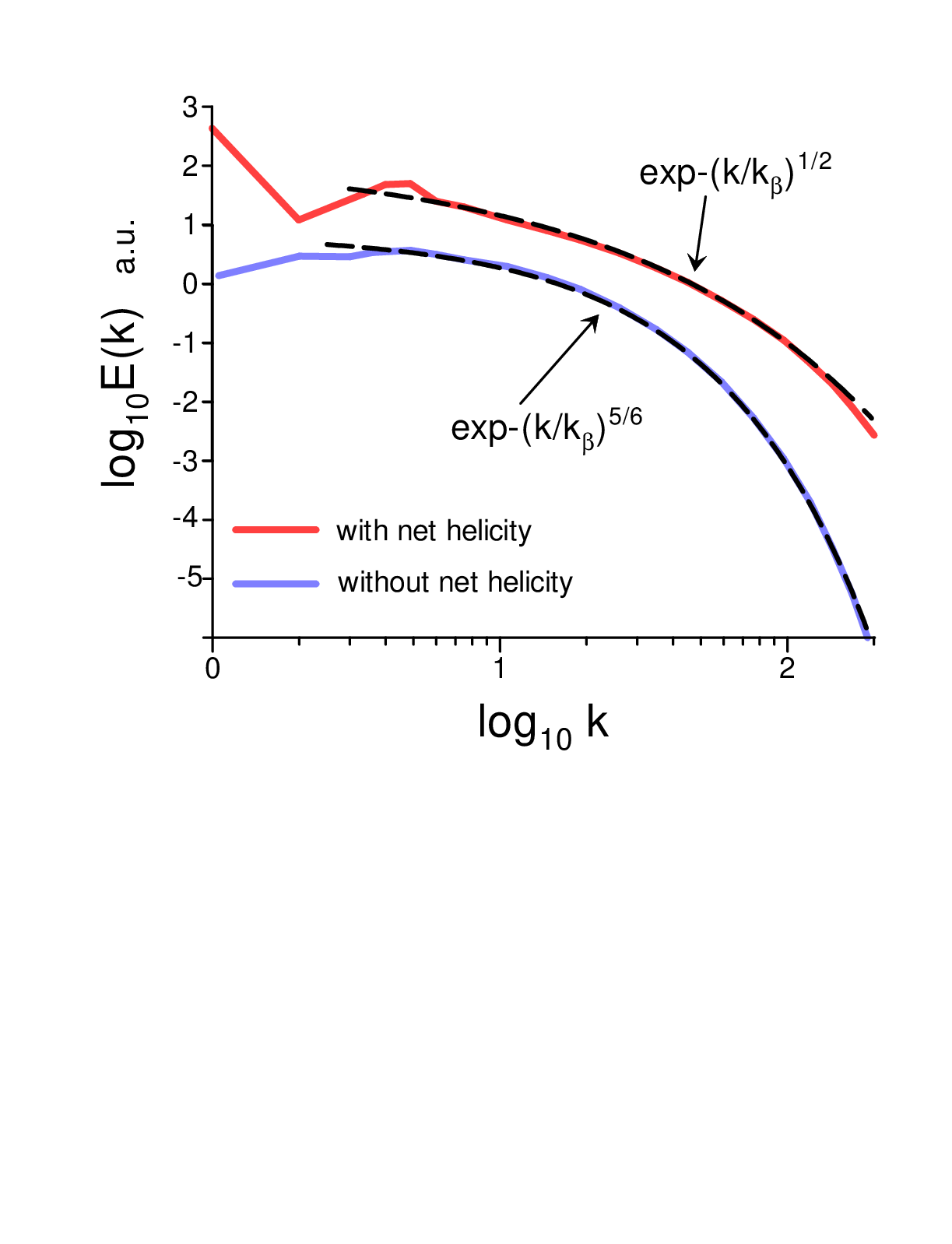} \vspace{-4.2cm}
\caption{Magnetic energy spectra generated by MHD dynamos at magnetic Prandtl number $Pr_m =1$ (direct numerical simulations). The spectra are vertically shifted for clarity.} 
\end{figure}
\begin{figure} \vspace{-0.5cm}\centering
\epsfig{width=.47\textwidth,file=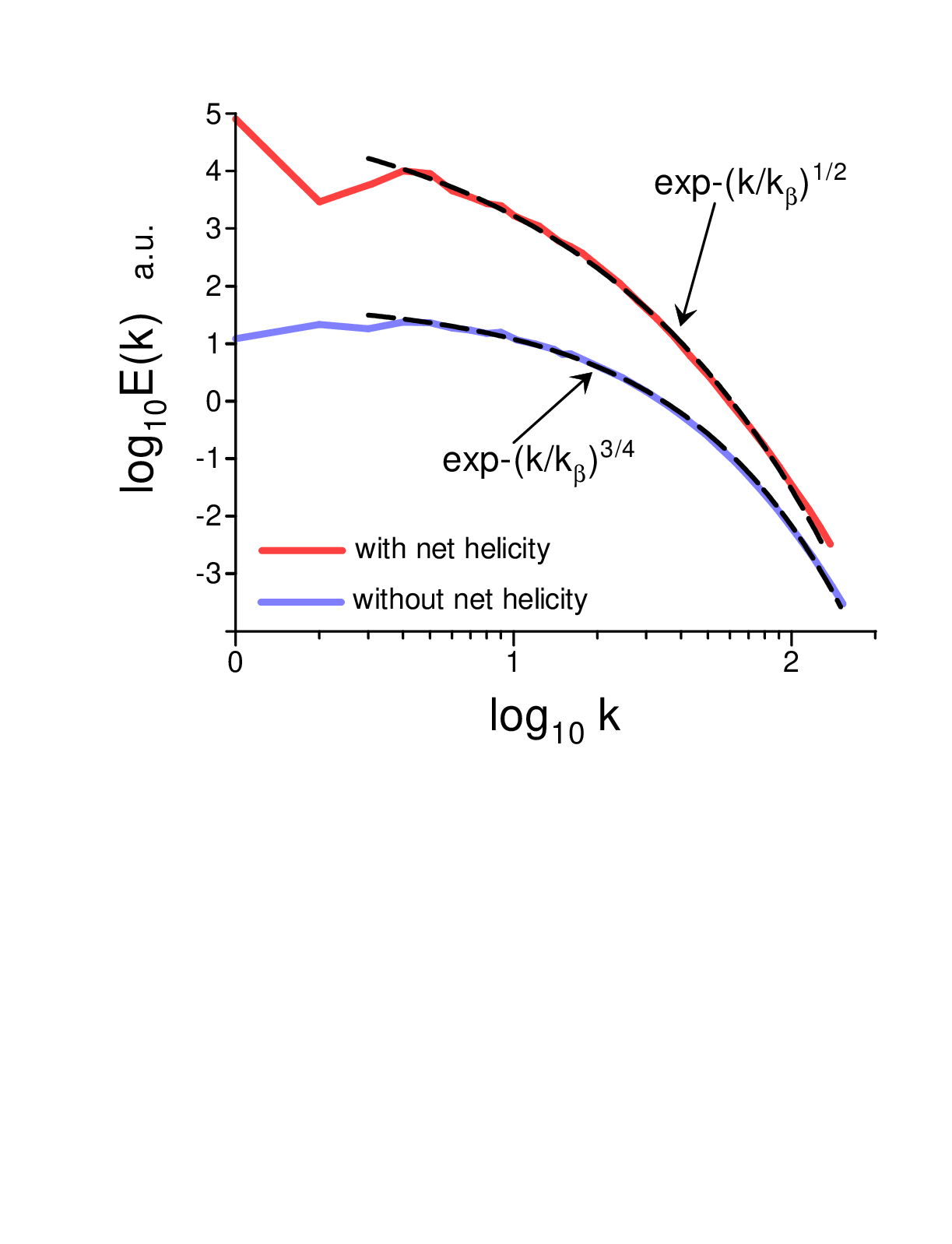} \vspace{-4.5cm}
\caption{Magnetic energy spectra generated by MHD dynamos at small magnetic Prandtl numbers $Pr_m =0.02$ and $Pr_m =0.01$ (direct numerical simulations). The spectra are vertically shifted for clarity.} 
\end{figure}
  In a special case of global (net) reflectional symmetry, the mean (net) magnetic helicity is equal to zero, while the point-wise magnetic helicity generally is not (due to the spontaneous breaking of the local reflectional symmetry). This is an intrinsic property of chaotic/turbulent flows. The blobs with non-zero kinetic/magnetic helicity can accompany the spontaneous symmetry breaking \cite{mt},\cite{kerr}-\cite{lt}. The magnetic blob is bounded by the magnetic surface with ${\bf b_n}\cdot {\bf n}=0$ (${\bf n}$ is a unit normal to the boundary of the magnetic blob). Sign-defined magnetic helicity of the magnetic blob is an ideal (adiabatic) invariant \cite{mt}. The magnetic blobs can be numbered and $H_j^{\pm}$ denotes the helicity of the magnetic blob with number $j$ and the blob's helicity sign `+'  or `-':
\begin{equation}
 H_j^{\pm} = \int_{V_j} ({\bf a} ({\bf x},t) \cdot  {\bf b} ({\bf x},t)) ~ d{\bf x} 
\end{equation}  
  
  Then we can consider the ideal adiabatic invariant 
\begin{equation}
{\rm I^{\pm}} = \lim_{V \rightarrow  \infty} \frac{1}{V} \sum_j H_{j}^{\pm}  
\end{equation} 
 The summation in Eq. (11) is over the blobs with a certain sign only (`+' or `-') and $V$ is the entire volume of the blobs over which the summation Eq. (11) was made.  \\

  The adiabatic invariant ${\rm I^{\pm}}$ Eq. (11) can be used instead of $h_m$ in the estimation Eq. (7) for the case of the local symmetry breaking
\begin{equation}
B_c \propto |{\rm I^{\pm}}|^{1/2} k_c^{1/2}   
\end{equation}
and the same spectrum Eq. (9) can be obtained for this case.

\subsection{Examples}

   One can recognize the spectrum Eq. (9) in the Figs. 1 and 2. 
   
   It should be noted that the DNS used to compute the spectra shown in Fig. 1 was globally nonhelical \cite{sch2} (see previous subsection C). 
   
   In this DNS the standard MHD equations in the Alfvenic units for incompressible fluid 
\bea
\partial_t {\bf u} + ({\bf u}\cdot \nabla) {\bf u} \!\!&=&\!\!  - \nabla P+ \left({\bf b}\cdot \nabla\right){\bf b} +\nu\nabla^2 {\bf u} + {\bf f} \\
\partial_t {\bf b}  + \left({\bf u}\cdot {\bf \nabla}\right){\bf b} \!\!&=&\!\! \left({\bf b}\cdot \nabla\right){\bf u} + \eta\nabla^2 \bf{b}  \\
\nabla \cdot {\bf u} &=& 0, ~~~   {\bf \nabla} \cdot {\bf b} =  0, 
\eea
were solved in a periodic spatial cube (here $P=p +\frac{{\bf b}^2}{2}$ and $p$ is the kinetic pressure normalized by the constant density). \\

  The dynamo DNS was initialized with seed random nonhelical magnetic fluctuations. The velocity forcing ${\bf f}$ was also nonhelical and random (white-noise-like). Therefore, the appearance of the spectrum Eq. (9) at $Re=622$ in this case should be related to the spontaneous breaking of local reflectional symmetry (further randomization - $\beta = 1/4$ for $Re= 1320$ also can be related to the symmetry breaking, see next Section).\\

    Figures 3 and 4 show the saturated magnetic energy spectra obtained with direct numerical simulations of the compressible isothermal gas isotropic motions in a periodic spatial cube (see for a more detailed description of the simulations setup in the papers Ref. \cite{bran1} -\cite{bran3}). The spectral data were taken from Figs. 1 and 4 of a paper Ref. \cite{bran4}. 
The upper spectra in these figures show the results of the simulations with net (global) helicity whereas the lower spectra show the simulations without net helicity. The upper spectrum in Fig. 3 was computed for the magnetic Prandtl number $Pr_m = 1$ and magnetic Reynolds number $Re_m = 450$ whereas the lower spectrum in Fig. 3 was computed for $Pr_m = 1$ and $Re_m = 600$. The upper spectrum in Fig. 4 was computed for $Pr_m = 0.01$ and $Re_m = 23$ ($Re=2300$) whereas the lower spectrum in Fig. 4 was computed for $Pr_m = 0.02$ and $Re_m = 230$ ($Re=11,500$). 

  As one can see a large-scale dynamo coexists with a small-scale dynamo for the (net) helical cases and only a small-scale dynamo exists in the (net) nonhelical cases in these direct numerical simulations. 
  
    The dashed curves indicate correspondence to the distributed chaos in these figures: correspondence to the helical spectrum Eq. (9) for the upper spectra and correspondence to the dissipative distributed chaos (see next Section) for the lower spectra.\\
  
     One can expect that the small-scale dynamos will play a more significant role at the large $Re_m$. The small-scale dynamo populates the chaotic/turbulent
fluid (plasma) with localized, highly intense structures and can affect the dynamic process faster than the large-scale dynamo \cite{rin}. The spontaneous breaking of local reflectional symmetry should also be more effective in this case. Therefore let us consider a small-scale dynamo at large $Re_m$. \\

\begin{figure} \vspace{-1.1cm}\centering
\epsfig{width=.48\textwidth,file=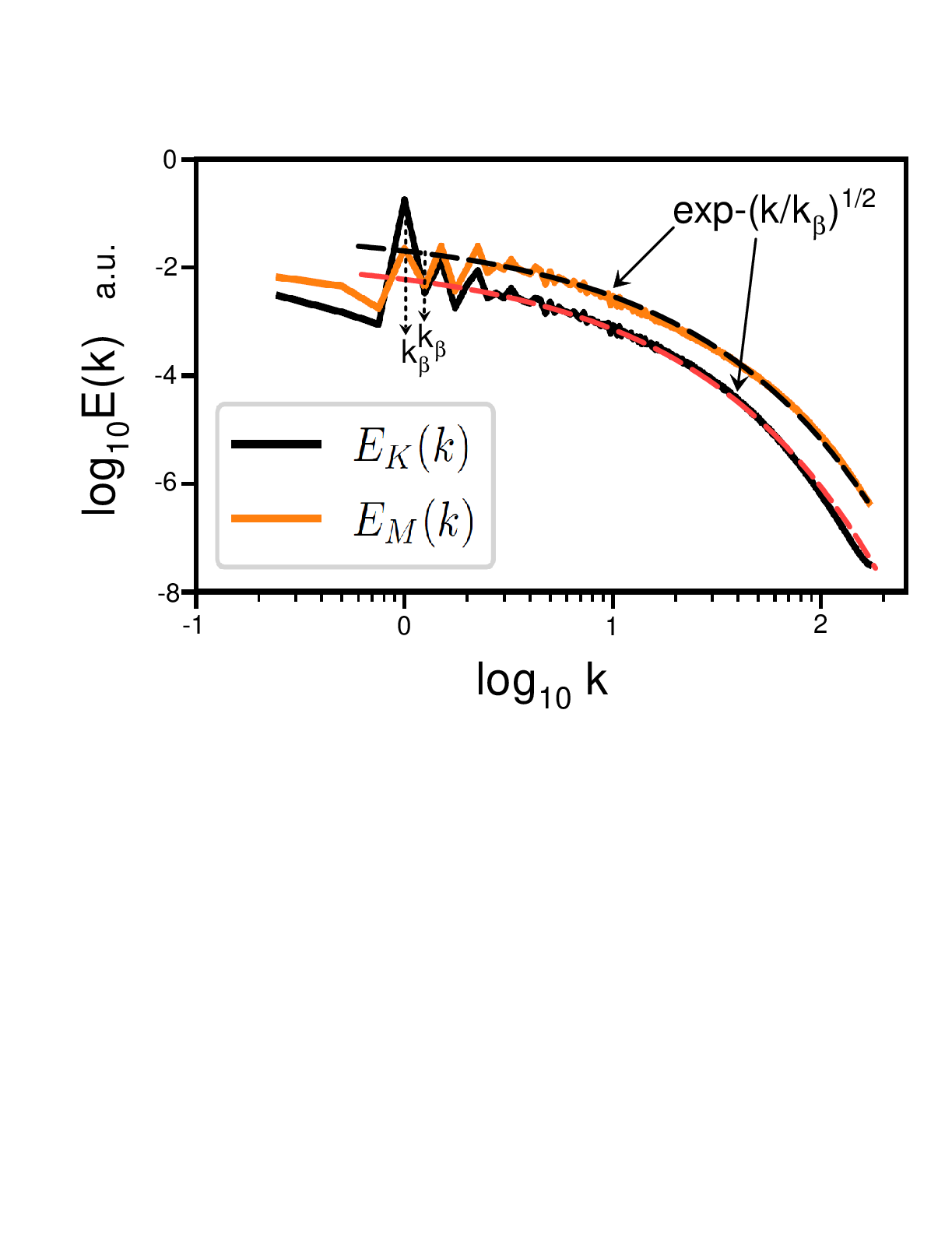} \vspace{-5.3cm}
\caption{Magnetic and kinetic energy spectra at $Re_m =2800$ and $Pr_m =4$ (direct numerical simulations without net kinetic and magnetic helicities). } 
\end{figure}

    Figure 5 shows magnetic (up) and kinetic (bottom) saturated energy spectra computed in a DNS reported in the recent paper \cite{rin}. The spectral data were taken from Fig. 5 of the Ref. \cite{rin}. The energy input ${\bf f}$ into the system Eqs. (13-15) was purely kinetic and nonhelical. The wavenumber in the Fig. 5 was normalized by the forcing wavenumber $k_f$. The magnetic Reynolds number $Re_m = 2800$ and the magnetic Prandtl number $Pr_m = 4$. Due to the global reflectional symmetry, the net helicity was negligible whereas the local helicities were strong (the spontaneous breaking of the local reflectional symmetry). The dashed curves indicate the helical energy spectra Eq. (9). \\
    
    There will be more examples of the helical spectrum Eq. (9) in the next sections.

 \section{Magneto-inertial range of scales}   

    In hydrodynamic turbulence presence of an inertial range of scales is expected for the high Reynolds numbers. In this range, the statistical characteristics of the motion depend on the kinetic energy dissipation rate $\varepsilon$ only \cite{my}. In magnetohydrodynamics, a magneto-inertial range of scales can be introduced.  In this range two parameters: the magnetic helicity dissipation rate $\varepsilon_h$ (or dissipation rate of its modification $I^{\pm}$ Eq. (11)) and the total energy dissipation rate $\varepsilon$ govern the magnetic field dynamics. An analogous situation with two governing parameters (kinetic dissipation rate and the passive scalar dissipation rate) was studied for the inertial-convective range in the Corrsin-Obukhov approach \cite{my} (see also Ref. \cite{bs1}). Let us, following this analogy, replace the estimates Eqs. (7) and (12) by the estimate
\begin{equation}
 B_c \propto \varepsilon_h^{1/2} ~\varepsilon^{-1/6}~k_c^{1/6}  
 \end{equation}
 for the magneto-inertial range dominated by magnetic helicity (for the modified magnetic helicity see Section VIIB).\\
 
 The estimates Eq. (7), (12), and (16) can be generalized as
\begin{equation}
 B_c \propto k_c^{\alpha}   
 \end{equation}
  
   Let us look for the spectrum of the distributed chaos as a stretched exponential (see Introduction and Eqs. (6) and (9))
\begin{equation}
E(k) \propto \int_0^{\infty} P(k_c) \exp -(k/k_c)dk_c \propto \exp-(k/k_{\beta})^{\beta} 
\end{equation}    

  The Eq. (18) can be used to estimate the probability distribution $P(k_c)$ for large $k_c$ \cite{jon}
\begin{equation}
P(k_c) \propto k_c^{-1 + \beta/[2(1-\beta)]}~\exp(-\gamma k_c^{\beta/(1-\beta)}) 
\end{equation}     
  
   A relationship between the exponents $\beta$ and $\alpha$ can be readily obtained (using some algebra) from the Eqs. (17) and (19) for the half-normally distributed positive variable $B_c$ 
\begin{equation}
\beta = \frac{2\alpha}{1+2\alpha}  
\end{equation}

   Since for the magneto-inertial range (dominated by the magnetic helicity) $\alpha =1/6$ (see Eq. (16)) corresponding magnetic energy spectrum can be estimated as
\begin{equation}
 E(k) \propto \exp-(k/k_{\beta})^{1/4}  
 \end{equation}
  
   The spectrum Eq. (21) can be recognized in the Figs. (1-2) and can be considered as a precursor of the hard turbulence. \\
   
   Figure 6 shows magnetic energy spectra computed in a recent paper Ref. \cite{warne} using direct numerical simulations of isothermal forced turbulence. The paper Ref. \cite{warne} numerically studied the small-scale dynamo at rather low Prandtl numbers and considerably large Reynolds numbers. The spectral data were taken from Fig. 4 of the Ref. \cite{warne}. The boundary conditions were periodic and the small-scale dynamo was initialized by a seed magnetic field with weak Gaussian noise.
   
   The dashed curves indicate the magnetic energy spectrum  Eq. (2) (deterministic chaos) for $Re = 7,958$, and $Pr_m = 0.05$, and the magnetic energy spectrum Eq. (21) for  $Re = 32,930$ and $Pr_m = 0.005$ (distributed chaos in the magneto-inertial range of scales).\\ 
   
   More examples of the spectrum Eq. (21) will be given in the next sections.

 \section{Dissipative distributed chaos}
 
\subsection{Dissipative MHD invariants}

\begin{figure} \vspace{-0.8cm}\centering
\epsfig{width=.45\textwidth,file=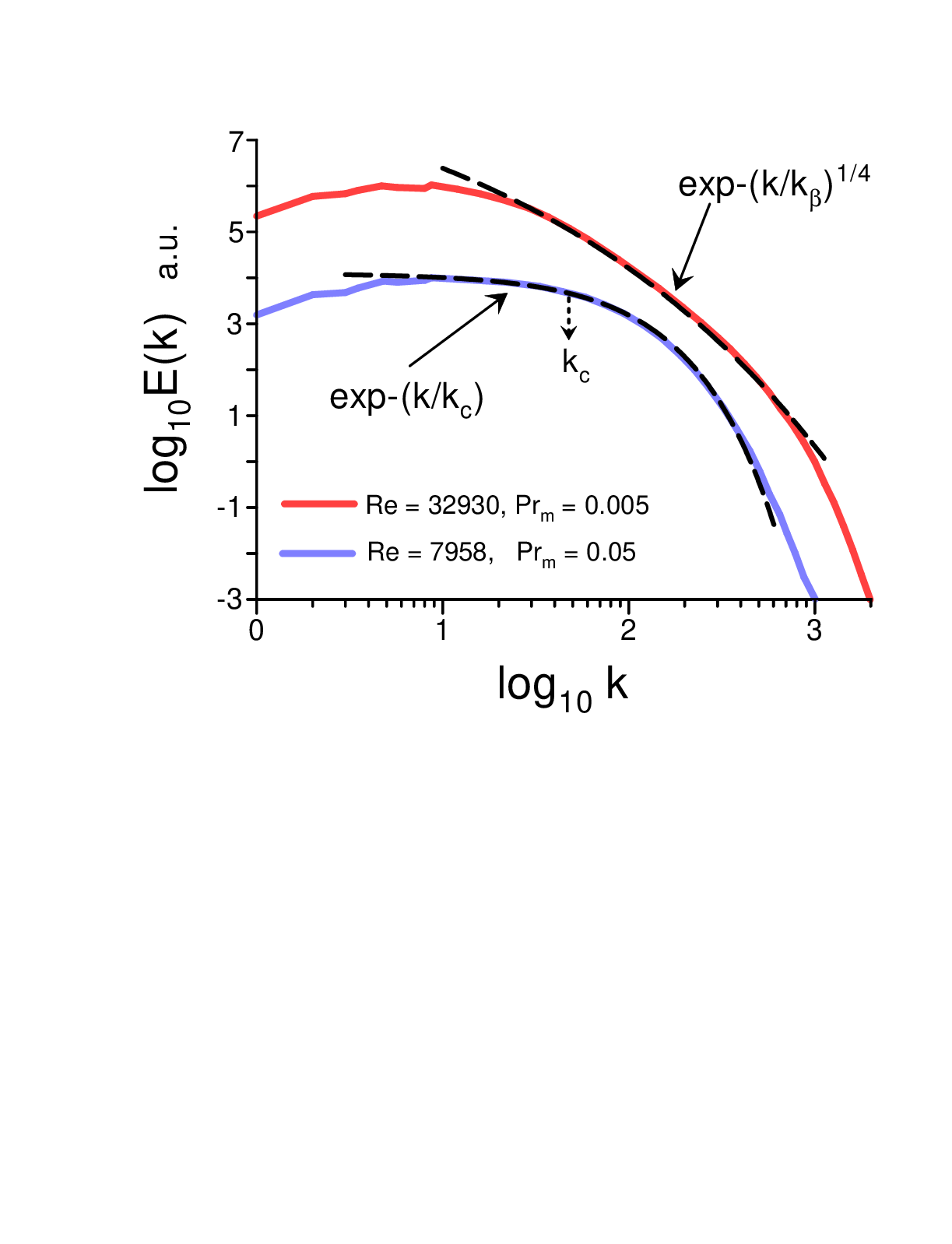} \vspace{-4.7cm}
\caption{Magnetic energy spectra at $Re = 7,958$, and $Pr_m = 0.05$ (bottom), and at $Re = 32,930$ and $Pr_m = 0.005$ (up). The direct numerical simulations were performed without net kinetic and magnetic helicities. } 
\end{figure}

 Dissipative invariants for hydrodynamics (for the Navier-Stokes equations) were introduced by Loitsianskii \cite{my} (Loitsianskii integral), and Birkhoff \cite{bir} and Saffman (Birkhoff-Saffman integral) \cite{saf}. Chandrasekhar extended this notion on the dissipative magnetohydrodynamics \cite{chan}.  \\
 
   Originally the dissipative invariants were used to estimate the energy spectra at small wavenumbers \cite{my},\cite{saf}. Therefore, the restrictions related to isotropy and homogeneity were rather strict as well as the possible extension of the notion of the dissipative invariants on magnetohydrodynamics. However, if one applies these invariants to the larger wavenumbers (as it will be done below in the frames of the distributed chaos notion) these restrictions can be eased to the local isotropy and homogeneity and the adiabatic invariance (i.e. a slow change with time in comparison to the time scales characteristic to the distributed chaos range).  \\
   
   The MHD K\'{a}rm\'{a}n-Howarth equation in terms of Els\"{a}sser variables ${\bf z}^{\pm}={\bf v}\pm {\bf b}$ can be written as  \cite{pp}
\begin{equation}
\frac{\partial\langle z_L^{\pm}z_L^{\pm '}\rangle}{\partial t} = \left(\frac{\partial}{\partial r}+\frac{4}{r}\right)C^{\pm}_{LLL}(r)+2\left(\frac{\partial^2}{\partial r^2} 
+ \frac{4}{r} \right) D_{LL}  
\end{equation}
where the subscript $L$ denotes projections of ${\bf z}^{\pm}$ on ${\bf r}$ (longitudinal, i.e. $z_L^{\pm} = {\bf z}^{\pm} \cdot {\bf r}/r$), ${\bf z}^{\pm '} = {\bf z}^{\pm}({\bf x}+{\bf r}, t)$, $D_{LL}(r) = [\nu_+\langle z_L^{\pm}z_L^{\pm '}\rangle+\nu_-\langle z_L^{\pm}z_L^{\mp '}\rangle]$, $C^{\pm}_{LLL}(r)=\langle z_L^{\pm}z_L^{\mp}z_L^{\pm '} \rangle$, $\nu_{\pm} = \nu \pm \eta$.

  Multiplying both sides of the K\'{a}rm\'{a}n-Howarth Eq. (22) by $r^4$ and then integrating them on $r$ from 0 to $R$ we obtain
\begin{equation}
{\frac{\partial\int\limits_{0}^{R} r^4\langle z_L^{\pm}z_L^{\pm '}\rangle dr}{\partial t}}=R^4C^{\pm}_{LLL}(R)\left. +2 R^4\frac{\partial D_{LL}}{\partial r}\right\arrowvert_{r=R}  
\end{equation}
If the terms
\begin{equation}
C^{\pm}_{LLL}(R)\left.~~~~\rm{and}~~~~\frac{\partial D_{LL}}{\partial r}\right\arrowvert_{r=R} \rightarrow 0
\end{equation}
fast enough when $R \rightarrow \infty$, then 
\begin{equation}
\lim_{r\rightarrow R} {\frac{\partial\int\limits_{0}^{R} r^4\langle z_L^{\pm}z_L^{\pm '}\rangle dr}{\partial t}} =0   
\end{equation}
 and, as a consequence,
\begin{equation}
\int r^2 \langle z_L^{\pm}z_L^{\pm '}\rangle d{\bf r} = \rm{constant}  
\end{equation}
Returning to the velocity and magnetic fields we obtain
\begin{equation}
 \int r^2[\langle v_Lv_L'\rangle + \langle b_L  b_L'\rangle]~ d{\bf r} = \rm{constant}  
\end{equation} 

  It is shown in the paper \cite{chan} that 
\begin{equation}
\int r^2 \langle v_Lv_L'\rangle~ d{\bf r} = \rm{constant}  
\end{equation}
Then it follows from the Eqs. (26-27) 
\begin{equation}
\mathcal{L}_b = \int r^2\langle b_L  b_L'\rangle~ d{\bf r} = \rm{constant}  
\end{equation} 
  
  The integral $\mathcal{L}_b$ can be considered as a magnetic analogy of the Loitsianskii (adiabatic) invariant. \\
  
  Analogously, using another form of the K\'{a}rm\'{a}n-Howarth equation suggested in the paper Ref. \cite{wan}, it can be shown that the magnetic analogy of the Birkhoff-Saffman integral is also a dissipative (adiabatic) invariant \cite{ber4}
\begin{equation}
 \mathcal{S}_b = \int \langle {\bf b}  \cdot{\bf b}'\rangle~ d{\bf r} = \rm{constant}  
\end{equation}  
  
 \subsection{Dissipative distributed chaos}   
    
    In a range of scales dominated by the magnetic Birkhoff-Saffman invariant Eq. (29), the estimate Eq. (7) should be replaced by the estimate 
\begin{equation}
 B_c \propto \mathcal{S}_b^{1/2} k_c^{3/2}   
\end{equation}  
 i.e. $\alpha =3/2$ in his case. Using the relationship Eq. (20) we obtain $\beta =3/4$, that corresponds to the magnetic energy spectrum 
\begin{equation}
E(k)    \propto \exp-(k/k_{\beta})^{3/4}. 
\end{equation}

   Analogously, in a range of scales dominated by the magnetic Loitsianskii invariant Eq. (28), the estimate Eq. (7) should be replaced by the estimate 
\begin{equation}
 B_c \propto \mathcal{L}_b^{1/2} k_c^{5/2}   
\end{equation}    
i.e. $\alpha =5/2$ in his case. Using the relationship Eq. (20) we obtain $\beta =5/6$, that corresponds to the magnetic energy spectrum   
\begin{equation}
E(k)    \propto \exp-(k/k_{\beta})^{5/6}.  
\end{equation}

 One can recognize the spectra Eqs. (31) and (33) in the Figs. (1-4). \\

   Figure 7 shows the magnetic energy spectra for different values of Reynolds number  (the spectral data were taken from Fig. 2 of a paper Ref. \cite{hbd}). The DNS of a dynamo reported in the Ref. \cite{hbd} was performed in nonhelical subsonic motion of isothermal gas with the constant sound speed in a periodic cubic box with a large-scale random hydrodynamic forcing at $Pr_m =1$ (the variety of the Reynolds numbers is shown in Fig. 7). 
   
   The dashed curves indicate the best fit by the spectrum Eq. (2) (deterministic chaos at $Re =190$) and by the spectra Eq. (33) and Eq. (31) (dissipative distributed chaos at $Re =420$ and $Re = 960$, respectively). The randomness becomes stronger with increasing Reynolds number (cf. Fig. 1). \\
 
\begin{figure} \vspace{-1cm}\centering \hspace{-1cm}
\epsfig{width=.50\textwidth,file=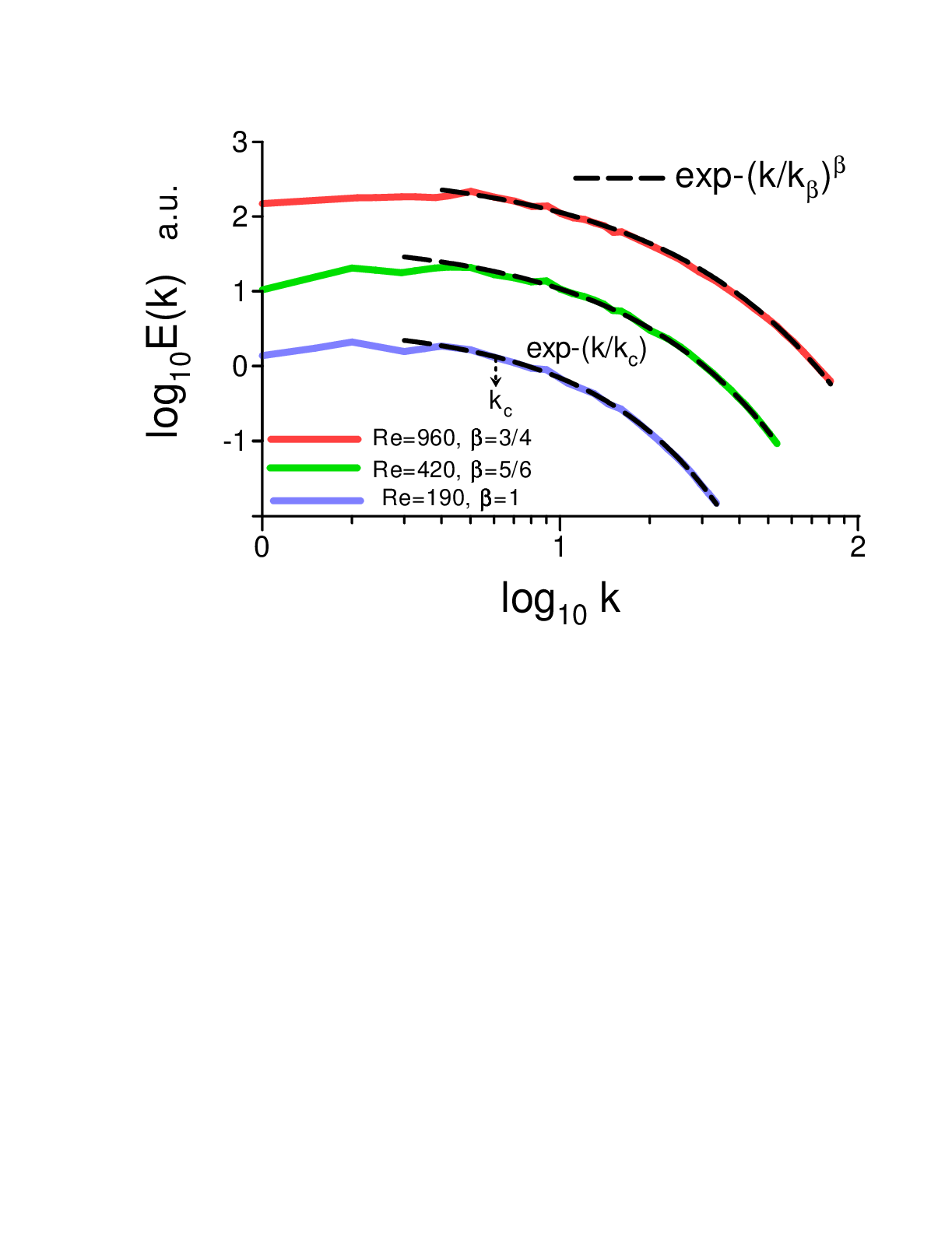} \vspace{-5.9cm}
\caption{Magnetic energy spectra for different values of Reynolds number and $Pr_m =1$. The direct numerical simulations were performed without net kinetic and magnetic helicities. The spectra are vertically shifted for clarity.} 
\end{figure}
\begin{figure} \vspace{-0.5cm}\centering
\epsfig{width=.45\textwidth,file=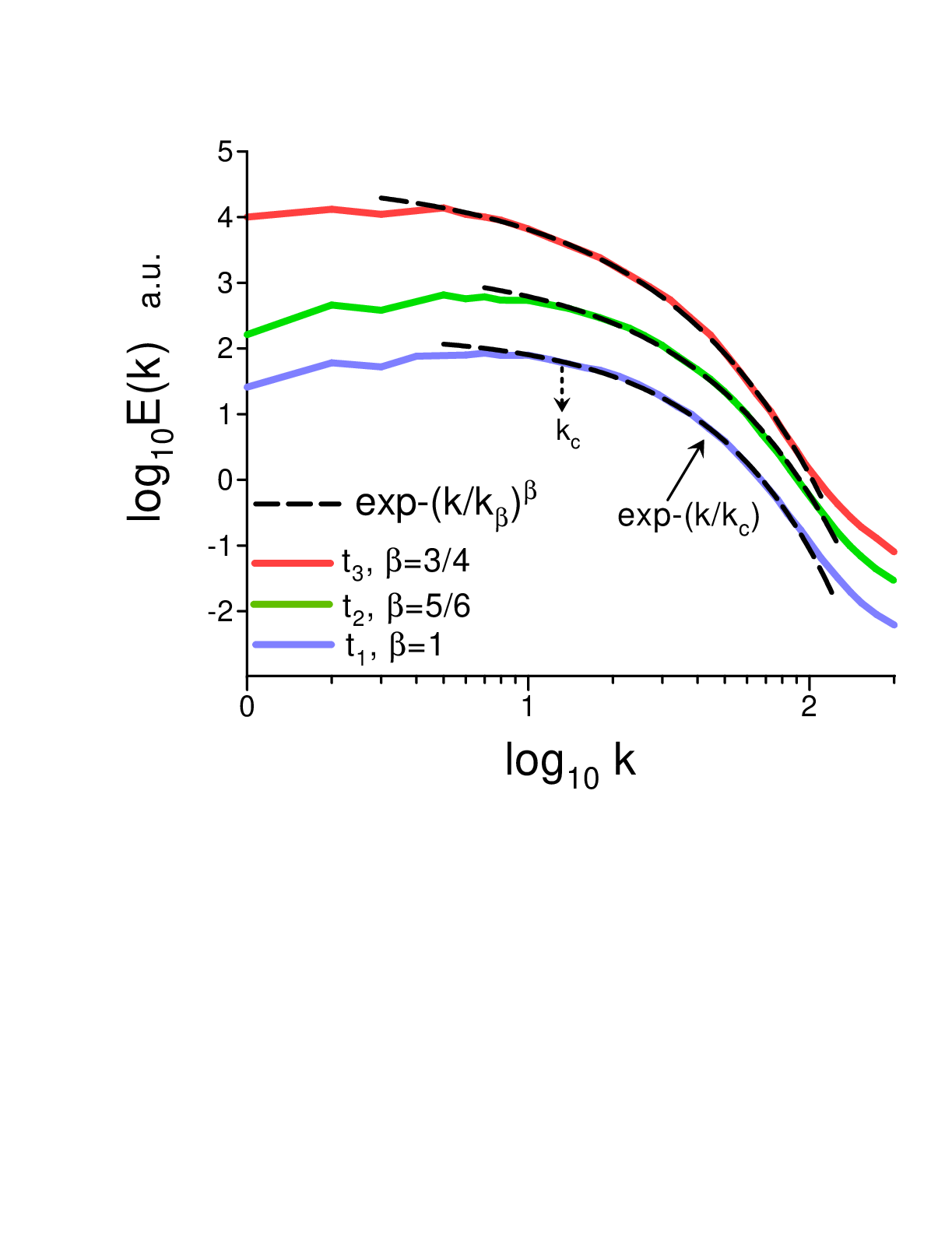} \vspace{-4cm}
\caption{Magnetic energy spectra for different times of evolution ($t_1 < t_2 < t_3$) of a direct numerically simulated small-scale dynamo ($Re = 1250$, and $Pr_m =1$) . The direct numerical simulations were performed without net kinetic and magnetic helicities. The spectra are vertically shifted for clarity. } 
\end{figure}
 
   Figure 8 shows the magnetic energy spectra for different times of evolution ($t_1 < t_2 < t_3$) of a direct numerically simulated small-scale dynamo (the spectral data were taken from Fig. 2 of a recent paper Ref. \cite{sbs}). In this DNS a non-helical, statistically isotropic and homogenous, small-scale dynamo in a randomly forced (a stochastic Ornstein–Uhlenbeck process) gas motion was performed in a periodic spatial cube. The considered here transonic case corresponds to the Mach number $M =1.1$, $Re = 1250$, and $Pr_m =1$ (the transonic motion consists of the regions of both supersonic and subsonic motions). \\
   
   The dashed curves indicate the best fit by the spectrum Eq. (2) (deterministic chaos at $t_1$) and by the spectra Eq. (33) and Eq. (31) (dissipative distributed chaos at $t_2$ and $t_3$, respectively). The randomness becomes stronger with increasing time of the dynamo evolution (cf. Fig. 2). \\
   
   More examples of the spectra Eqs. (31) and (33) will be given in the next sections.

\section{Hall magnetohydrodynamics }
   
 In the ideal Hall MHD (unlike the standard magnetohydrodynamics) ions are not tied closely to the magnetic field because of the ions' inertia. Electrons (which are still tied to the magnetic field) and the ions decouple at the ions' inertial length $d_i$. 

  The equations taking into account this phenomenon can be written in the form
\bea
\partial_t {\bf u} + ({\bf u}\cdot \nabla) {\bf u}  \!\!&=&\!\!  -\nabla P + \left({\bf b}\cdot \nabla\right){\bf b} +\nu\nabla^2 {\bf u} + {\bf f} \\
\partial_t {\bf b} + \left({\bf u}\cdot {\bf \nabla}\right){\bf b} \!\!&=&\!\! \left({\bf b}\cdot \nabla\right){\bf u} + \eta\nabla^2 \bf{b} \nonumber \\
 && \!\!\!\!\!\!\!\!\! \!\!\!\!\!\!\!\!\!\!\!\!\!\!\!\!\!\!\!   + d_i\left({\bf j}\cdot {\bf \nabla}\right){\bf b} -d_i\left({\bf b}\cdot \nabla\right){\bf j} \\
\nabla \cdot {\bf u}  &=& 0, ~~ {\bf \nabla} \cdot {\bf b}  = 0 
\eea
The last two terms of Eq. (35) correspond to the Hall effect, where  ${\bf j}= \nabla \times {\bf b}$ (cf. Eqs. (13-15) for standard magnetohydrodynamics) . \\

   Magnetic helicity is an ideal invariant for the Hall magnetohydrodynamics \cite{sch1},\cite{pm}) and the K\'{a}rm\'{a}n-Howarth equation can be written for the Hall magnetohydrodynamics \cite{galt}. Therefore, the spectral laws Eqs. (9), (21), (31), and (33) can be obtained for the Hall magnetohydrodynamics as well .\\

   In a paper Ref. \cite{map} results of a DNS of a Hall MHD dynamo were reported. The DNS was performed in a periodic spatial cube with the size $L_0$ and the helical large-scale hydrodynamic forcing was chosen as an ABC flow. A random small-scale seed magnetic field was introduced into the system after the flow reached a statistically steady state. \\
   
   Figure 9 shows the time evolution of the magnetic energy spectrum at $\varepsilon_H = d_i/L_0 =0.1$ (the spectral data were taken from Fig. 2b of the paper Ref. \cite{map}). The dashed curve at $t=5$ (the best fit) indicates the exponential spectrum Eq. (2) (i.e. the deterministic chaos). The dashed curves at $t=15$ and $t=30$ indicate the stretched exponential spectra Eq. (33) and Eq. (31) respectfully, i.e. the dissipative distributed chaos dominated by the magnetic Loitsianskii and Birkhoff -Saffman adiabatic invariants. At the later time $t=45$ of the dynamo development, we can observe the helical spectrum Eq. (9).  \\
   
   One can see that the randomness increases with the development of the Hall dynamo (cf. previous sections). \\

\section{Geodynamo}

\begin{figure} \vspace{-0.5cm}\centering
\epsfig{width=.45\textwidth,file=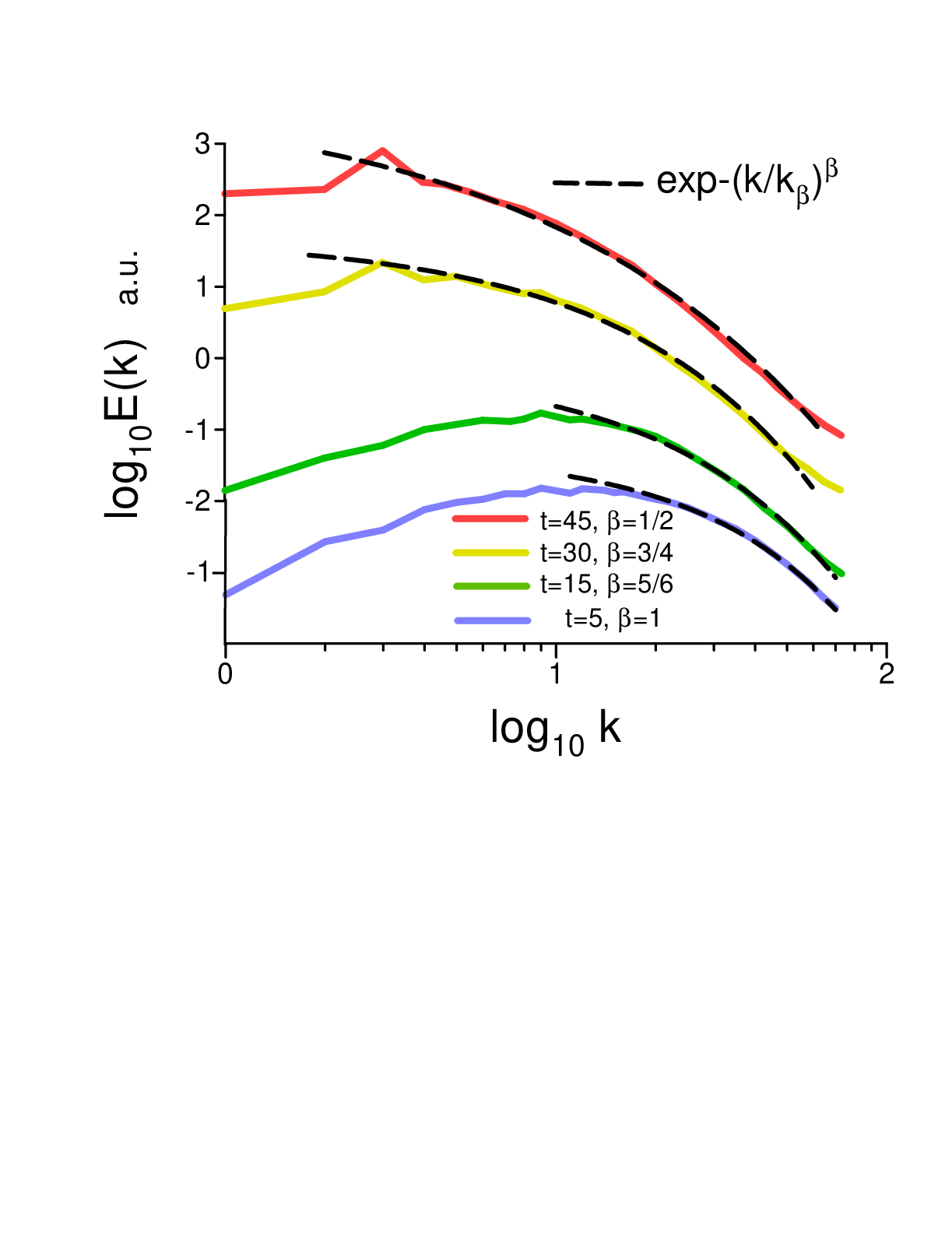} \vspace{-4.3cm}
\caption{Magnetic energy spectra for different times of evolution of a direct numerically simulated small-scale helical Hall MHD dynamo. The spectra are vertically shifted for clarity. } 
\end{figure}
 \begin{figure} \vspace{-0.5cm}\centering \hspace{-1cm}
\epsfig{width=0.45\textwidth,file=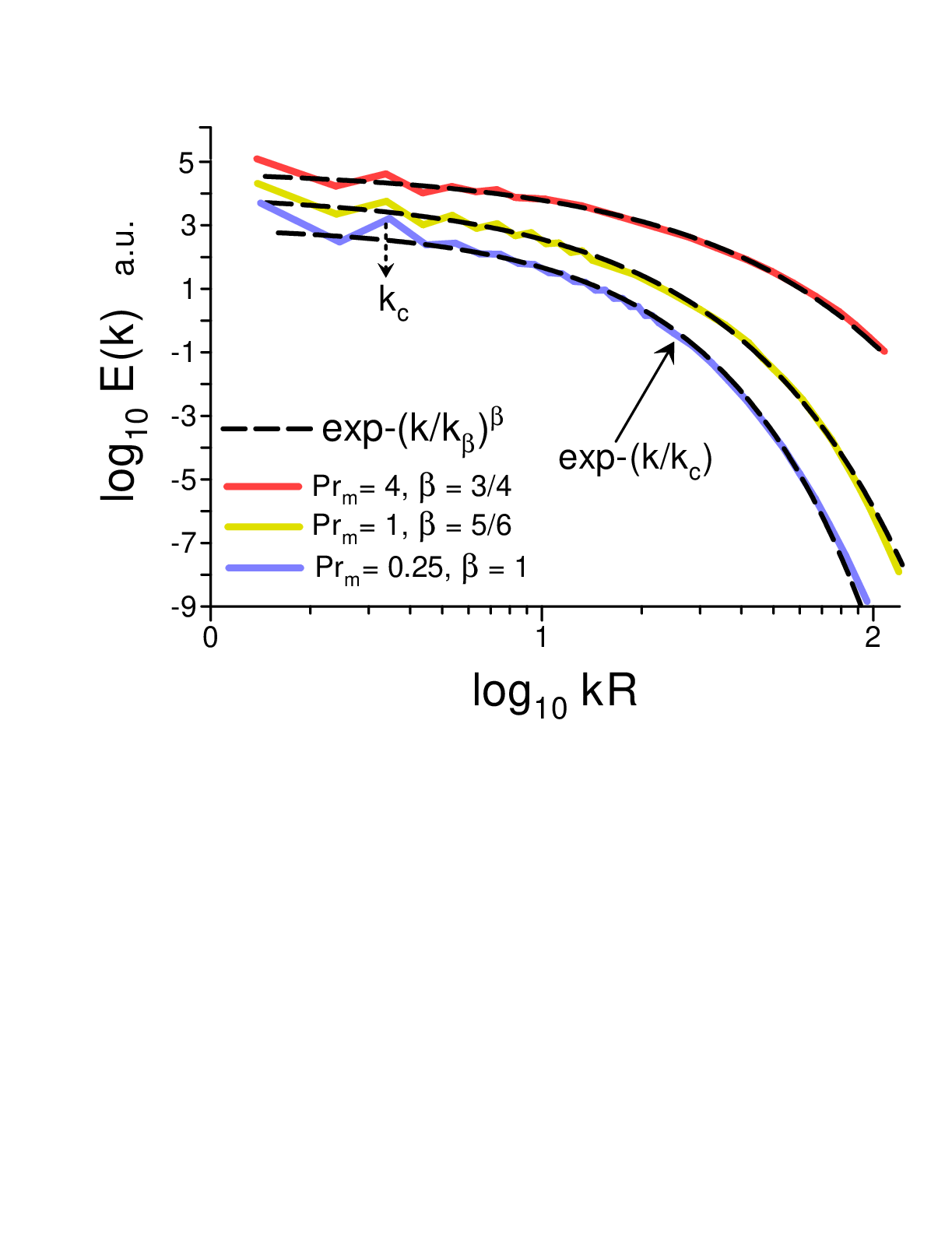} \vspace{-4.55cm}
\caption{Magnetic energy spectrum $E(k)$ generated by a chaotic magnetohydrodynamic dynamo in a conducting spherical Couette flow at $Pr_m = 0.25, 1, 4$ (a direct numerical simulation). } 
\end{figure}

  It is now believed that the magnetic field of Earth is generated by a flow of liquid metal in the Earth's core (the geodynamo). The measurements inside the core are impossible while the systematic good quality measurements of the spatial distribution of the magnetic field on the Earth's surface are still rare. Therefore, direct numerical simulations can be rather useful. \\
  
  Figure 10 shows the magnetic energy spectra $E(k)$ (against the wavenumber $k$) generated by a magnetohydrodynamic dynamo in a conducting spherical Couette flow at the magnetic Prandtl numbers $Pr_m = \nu/\eta =0.25, 1, 4$ (here $\nu$ is the kinematic viscosity and $\eta$ is the magnetic diffusivity of the fluid) and the Ekman number $E = \nu/( \Omega r_o^2) = 10^{-3}$ (here $\Omega$ is the rotational rate of the outer sphere and $r_o$ is its radius). This is a flow between two concentric spheres with a differential rotation relevant to the Earth's outer liquid core. To simulate the situation in the Earth's core the aspect ratio $r_i/r_o = 0.35$ (here $r_i$ is the radius of the inner core, see for more details \cite{gc}). 
   
   The spectral data were taken from Fig. 13 of the Ref. \citep{gc}. The spectra were time and $r$ averaged over the whole shell. $R$ is an effective radius ($r_i < R < r_o$). It should be noted that the spectral data were provided in  \citep{gc} as a dependence of the energy density spectra $E(l)$ on the spherical harmonic degree $l$. On a sphere of radius $R$ the wavenumber $k$ is related to $l$ as 
\bea
kR = \sqrt{l(l+1)} 
\eea
   
     The dashed curves indicate the best fit with the exponential Eq. (2) (the deterministic chaos), and stretched exponentials Eqs. (32) and (34) (the dissipative distributed chaos). Randomization increases with the $Pr_m$ in this case. The dotted vertical arrow indicates the position of the $k_c$. As one can see the small-scale dynamo coexists with the large-scale dynamo and the large-scale coherent structures determine the deterministic chaos (the position of the $k_cR$ coincides with the position of the spectral peak indicating the coherent structures).\\
     
\begin{figure} \vspace{-1cm}\centering \hspace{-1cm}
\epsfig{width=.495\textwidth,file=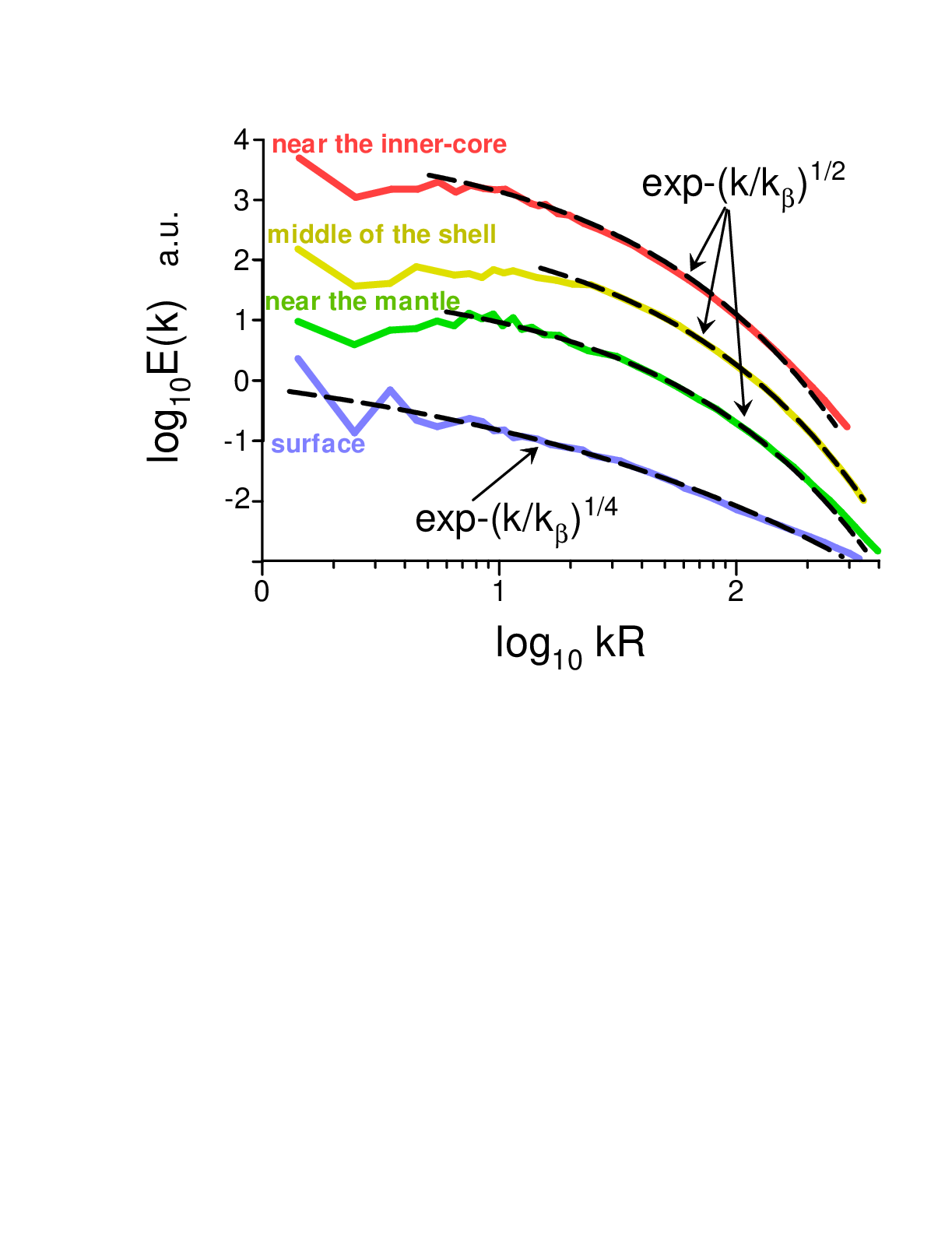} \vspace{-5.5cm}
\caption{Magnetic energy spectra at the spheres of different radii $R$. The spectra are vertically shifted for clarity. } 
\end{figure}
\begin{figure} \vspace{-1.15cm}\centering \hspace{-1cm}
\epsfig{width=.45\textwidth,file=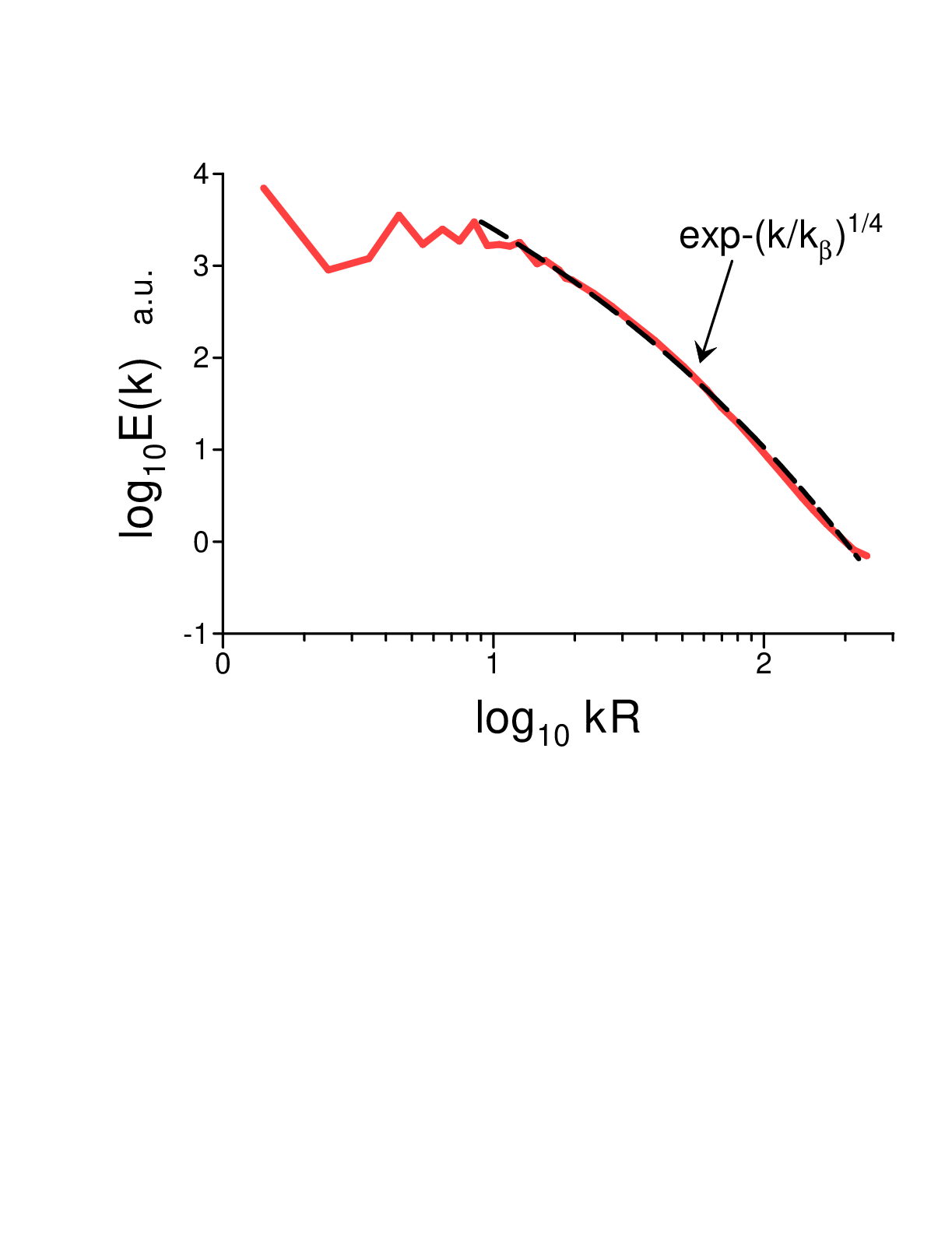} \vspace{-4.35cm}
\caption{Time-averaged magnetic energy spectrum at the Ekman number $E= 3\times 10^{-7}$ and $Pr_m =0.05$ (numerical simulation). } 
\end{figure}
     
   It is still very difficult to approach the values of the parameters characteristic for the real geodynamo (see, for instance, Table 1 in a recent paper Ref. \cite{sjna}).   In the Ref. \cite{sjna} a recent attempt to this end was made using maximal possibilities of the present computers. The authors of the DNS numerically solved the equations of the thermally (buoyancy) driven convection of an electrically conducting fluid with a chemical species concentration gradient (in the Boussinesq approximation with the Lorentz force) in a rotating spherical shell mimicking the liquid geodynamo confined between the mantle and solid inner core:
\bea
\partial_t {\bf u} +({\bf u}\cdot \nabla) {\bf u}  &=&-{\bf \nabla} p - \left(\frac{2}{E}\,{\bf e_z}\times {\bf u}\right) + \Delta {\bf u} +\nonumber \\
&&\!\!\!\!\!\!\!\!\!\!\!\!\!\!\!\!\!\!\!\!\!\!\!\!\!\!\!\!\!+ (\nabla \times {\bf b}) \times {\bf b} - \frac{Ra}{\beta}~C~\frac{D}{r_o}{\bf r}\\
\partial_t {\bf b} + \left({\bf u}\cdot {\bf \nabla}\right){\bf b} &=& \left({\bf b}\cdot \nabla\right){\bf u} + \frac{1}{Pr_m} \Delta {\bf b}  
\eea
\vspace{-0.5cm}
\bea
\partial_t C + {\bf u} \cdot {\bf \nabla} (C + C_0) &=& \frac{1}{Pr} \Delta C  \\
\nabla \cdot {\bf u} = 0, ~~~   {\bf \nabla} \cdot {\bf b} &=&  0  
\eea
where the gravity acceleration is proportional to the radius, $C({\bf r})$ is the scalar codensity field (for the chemical species concentration), $C_0(r)$ is the radial conductive codensity profile, $\beta$ is the codensity gradient, ${\bf e_z}$ is the unit vector in the direction of rotation, $E$ is the Ekman number, $Ra$ is the Rayleigh number, $D$ is the shell thickness, $r_0$ is the outer boundary radius.\\

  The boundary conditions for ${\bf u}$ are no-slip, the mantle and inner solid core are electrically insulating, and deviation of codensity from $C_0(r)$ on the boundaries has zero radial gradient (more details about choosing the  $C_0(r)$ and the initial conditions can be found in the Ref. \cite{sjna}).\\

   Figure 11 shows the magnetic energy spectra $E(k)$ against the wavenumber $k$ at the spheres of different radii $R$ (the spectral data were taken from Fig. 13 of the Ref. \cite{sjna} and transformed using the Eq. (41)). The dashed curves indicate the helical spectral law Eq. (9) near the inner-core, in the middle of the shell, and near the mantle, whereas at the surface the spectral law Eq. (21) indicates the magneto-inertial range of scales.\\
   
   Another example of a numerical simulation trying to approach the real conditions can be found in a recent paper Ref. \cite{sffj}. In this numerical simulation, the modeled liquid was enclosed in a rotating spherical shell between two radii $r_o$ and $r_i$ and the ratio $r_i/r_o=0.35$ (as for the Earth, cf the example of numerical simulation in Introduction). Both boundaries of the shell were taken no-slip and impermeable. The outer boundary was electrically insulating, and the inner solid core had the same conductivity as the liquid outer core. The inner core temperature was kept constant. Whereas the gradient of the temperature was kept constant and a uniform heat source was distributed throughout the outer core. The constant heat flux boundary conditions were applied at the core-mantle boundary. The dynamo was initiated and supported by the thermal convection under the strong influence of the rotation (the Ekman number $E = \nu/ \Omega (r_o-r_i)^2 = 3 \times 10^{-7}$, the magnetic Prandtl number $Pr_m =\nu/\eta =0.05$). \\
   
   Figure 12 shows the magnetic energy spectrum obtained at this numerical simulation (the spectral data were taken from Fig. 3b of the Ref. \cite{sffj}.  The spectrum was time and $r$ averaged over the whole shell, $R$ is an effective radius ($r_i < R < r_o$) and the wave number $k = \sqrt{l(l+1)}/R$ where $l$ is the spherical harmonic degree.
   
   The dashed curve indicates the spectral law Eq. (21), i.e. the magneto-inertial range for the small scales. One can see that the small-scale dynamo coexists with the large-scale one in this simulation as well.\\
   
   Figure 13 shows the geomagnetic power spectrum calculated using local subgrids from the World Magnetic Anomaly Map of the National Geophysics Data Center (WMAM) \cite{maus1}). The data were detrended and azimuthally averaged, and the mean field was subtracted by the author of the Ref. \cite{maus2} (the spectral data were taken from Fig. 4 of the Ref. \cite{maus2}). The land and ocean averages were considered separately (here $R$ is the Earth's mean radius). The dashed curve indicates the helical spectral law Eq. (9) for the ocean data, whereas at the land the spectral law Eq. (21) indicates the magneto-inertial range of scales. 
   
   The difference in the ocean and land spectra could be related to the electric conductivity of the mantle (see, for instance, \cite{shebalin2} and references therein) which can be different under ocean and land. This provides different local boundary conditions for the outer liquid core below the ocean and the land.\\
   
\begin{figure} \vspace{-0.9cm}\centering
\epsfig{width=.46\textwidth,file=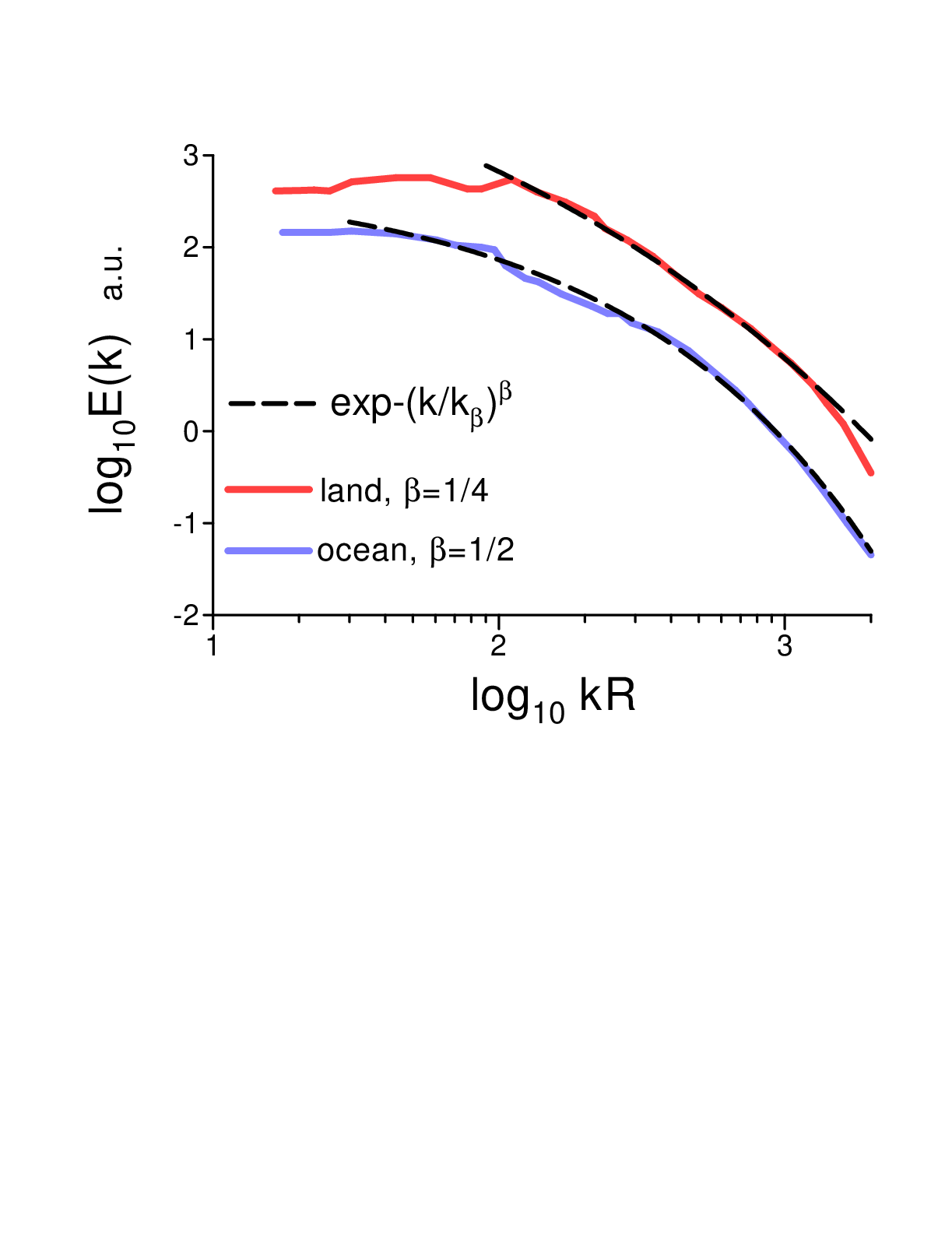} \vspace{-4.6cm}
\caption{Geomagnetic power spectrum calculated using local subgrids from World Magnetic Anomaly Map of the National Geophysics Data Center (WMAM). The spectra are vertically shifted for clarity. } 
\end{figure}
\begin{figure} \vspace{-0.5cm}\centering
\epsfig{width=.47\textwidth,file=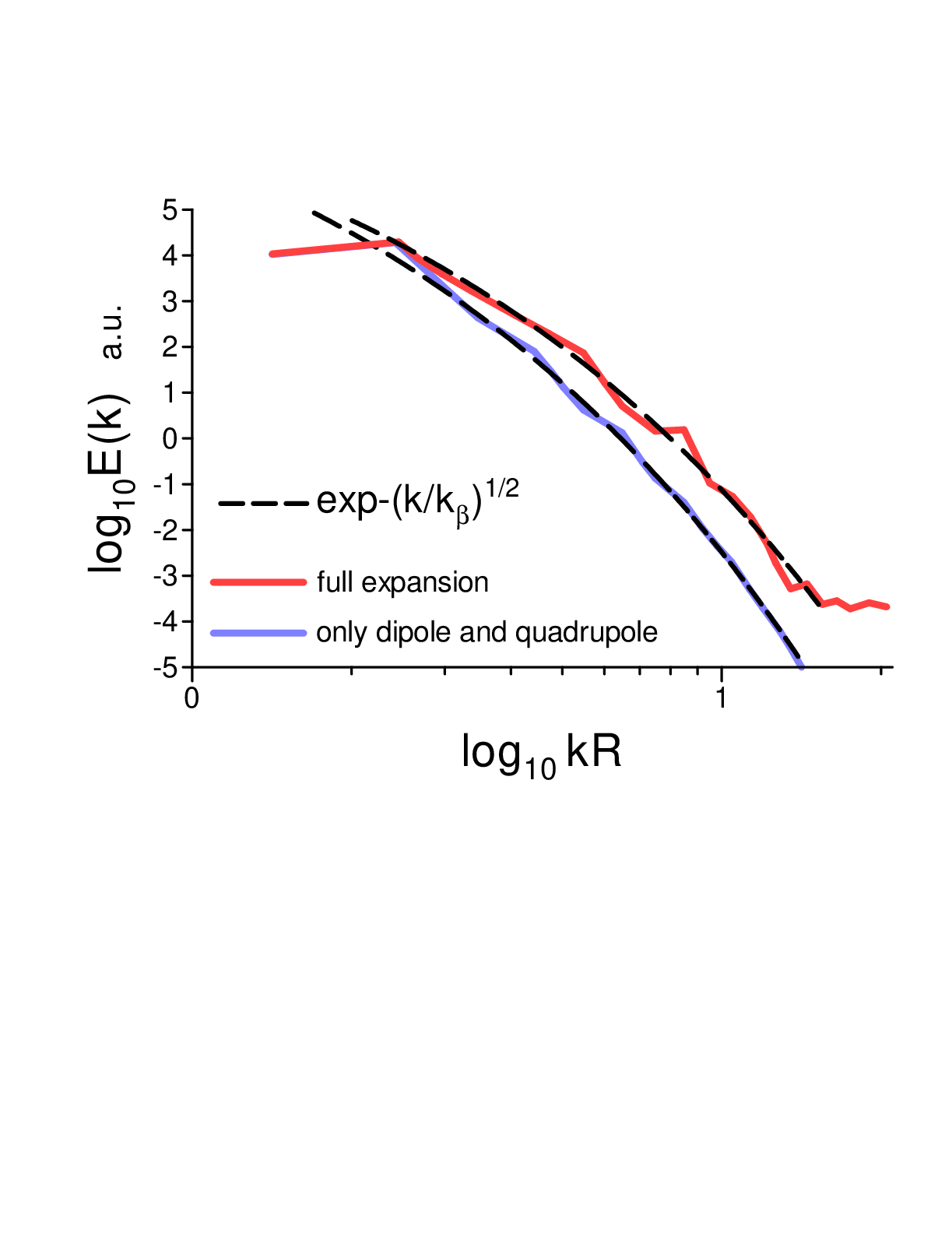} \vspace{-4.18cm}
\caption{Large-scale globally averaged total geomagnetic intensity spectrum. } 
\end{figure}
\begin{figure} \vspace{-0.9cm}\centering\hspace{-1cm}
\epsfig{width=.46\textwidth,file=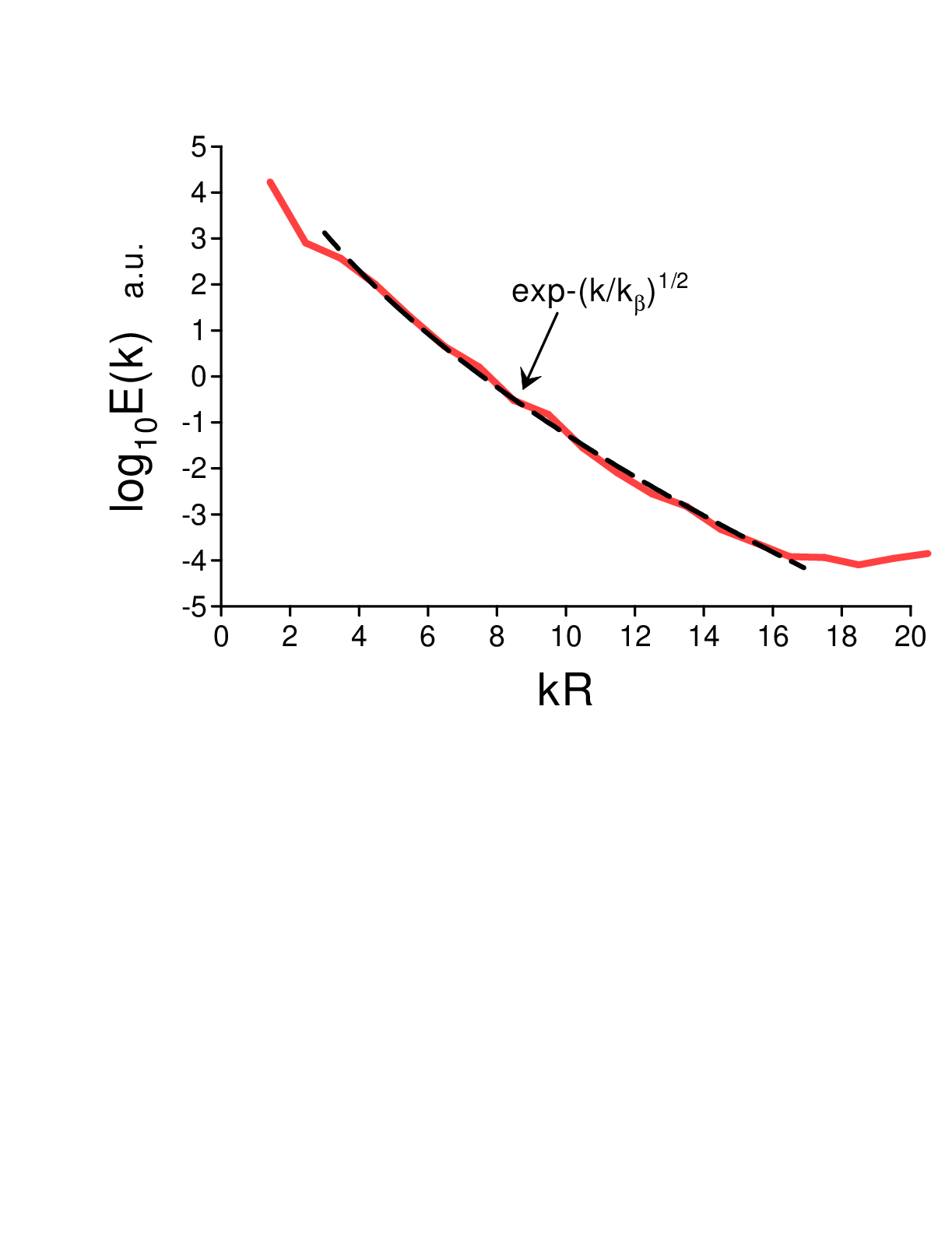} \vspace{-4.61cm}
\caption{Global geomagnetic energy spectrum at the Earth’s surface in the period 1999-2020.} 
\end{figure}
   In the paper Ref. \cite{maus2} a large-scale globally averaged total intensity spectrum computed using the spherical harmonic coefficients of the magnetic potential from a C89 global geomagnetic field model C89 \cite{cain} was also presented. It is interesting to compare the spectral data with the above-suggested spectra. Figure 14 shows the spectral data taken from Fig. 2 of the Ref. \cite{maus2}. The dashed curves indicate the helical spectrum Eq. (9)  (here $R$ is the Earth's radius).\\

     Figure 15 shows (in the semi-log scales) the spectral data taken from Fig. 2 of a recent paper \cite{finlay}). The spectral computations were produced using the CHAOS-7 model of the near-Earth magnetic field for the period 1999-2020 based on the observations made by the low-Earth orbit satellites CryoSat-2, Swarm, CHAMP, SAC-C, Orsted, and SAC-C, and on the ground observatory measurements. The dashed curve indicates the helical spectrum Eq. (9)  (here $R$ is the Earth's mean radius).

\section{Solar small-scale MHD dynamos}

  Both global (large-scale) and local (small-scale) chaotic/turbulent solar dynamos have been vigorously studied in the last decades (see, for instance, Refs.  \cite{cha}, \cite{ber2},\cite{remp} and references therein). It is believed that the chaotic/turbulent small-scale dynamo mechanisms are localized mainly in the near-surface solar layer. As it was already mentioned above the large- and small-scale dynamos can well coexist (see, for instance, Fig. 3) and both the quiet Sun and the highly active solar regions \cite{remp}\cite{ber3} can exhibit chaotic behavior. 
 
\subsection{Numerical simulations}

  Let us consider the results of two radiative MHD numerical simulations related to the small-scale solar dynamos. In these numerical simulations, the MHD equations for a compressible flow were extended by adding an equation for the total energy to take into account the radiative flux. \\
  
  In the first simulation \cite{kita}, both viscous and resistive dissipation were taken into account. Standard models of the lower solar atmosphere and the solar interior were taken for initial conditions. A weak seed uniform vertical magnetic field was added to a fully developed solar convection to initiate the quiet-Sun small-scale MHD dynamo. \\
  
  Figure 16 shows the magnetic energy spectra for a photosphere layer at different times of the dynamo evolution. The spectral data were taken from Fig. 3b of the Ref. \cite{kita} ($k_h$ corresponds to the horizontal wavenumber). The two earlier time spectra ($t=1$h and $t=2$h in the simulation terms) are well-fitted by the exponential spectrum Eq. (2) ($\beta =1$, deterministic chaos). Next spectrum ($t=4$h) can be fitted by the Eq. (33) ($\beta =5/6$, dissipative distributed chaos dominated by the magnetic Loitsianskii invariant), then the spectrum at $t=5$h can be fitted by Eq. (31) ($\beta =3/4$, dissipative distributed chaos dominated by the magnetic Birkhoff-Saffman invariant). \\
  
  The value of the parameter $\beta$ decreases with the time of the dynamo evolution, i.e. the randomness of the generated magnetic field increases with time.  \\ 
  
  Figure 17 shows the magnetic energy spectra also generated by a small-scale radiative MHD dynamo initiated by a seed uniform vertical magnetic field and averaged over a vertical range between $z = 0$ km and $z = -200$ km, and over time. The spectral data were taken from Fig. 3 of a recent paper Ref. \cite{riva}. The bottom curve corresponds to the case with a resistive dissipation whereas the top curve corresponds to the ideal case (without the resistive dissipation). For both cases, the grid resolution $h=6$ km. \\
  
    The dashed curves indicate the best fit by the Eq. (31) ($\beta =3/4$, dissipative distributed chaos dominated by the magnetic Birkhoff-Saffman invariant) for the dissipative case, and by the Eq. (9) ($\beta =1/2$, helical distributed chaos) for the ideal (non-dissipative case). 
    
 \subsection{Solar observations}  
\begin{figure} \vspace{-0.8cm}\centering \hspace{-0.8cm}
\epsfig{width=.51\textwidth,file=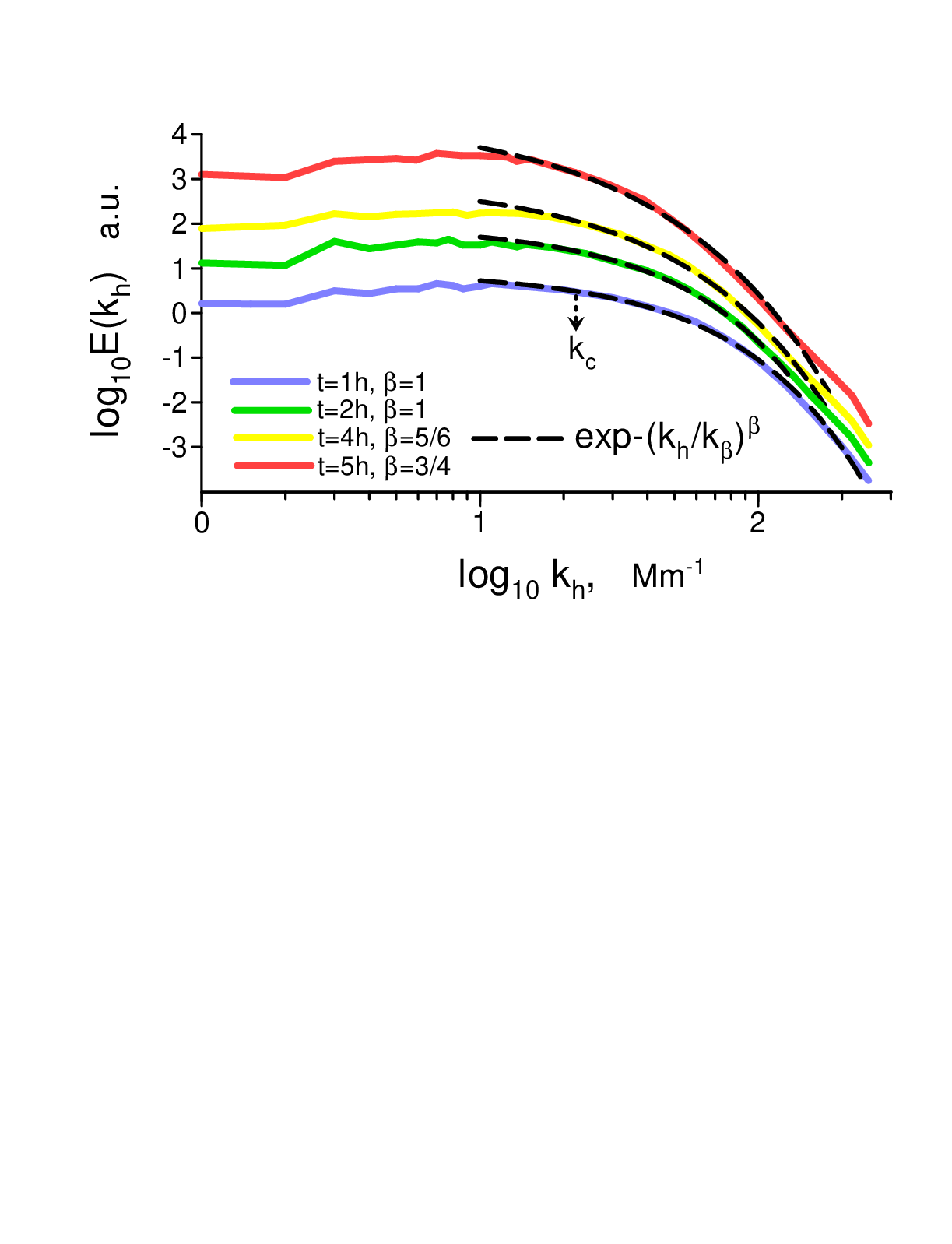} \vspace{-6.1cm}
\caption{Magnetic energy spectra for a photospheric layer at different times of the dynamo evolution (DNS). } 
\end{figure}
\begin{figure} \vspace{-0.45cm}\centering \hspace{-0.8cm}
\epsfig{width=.51\textwidth,file=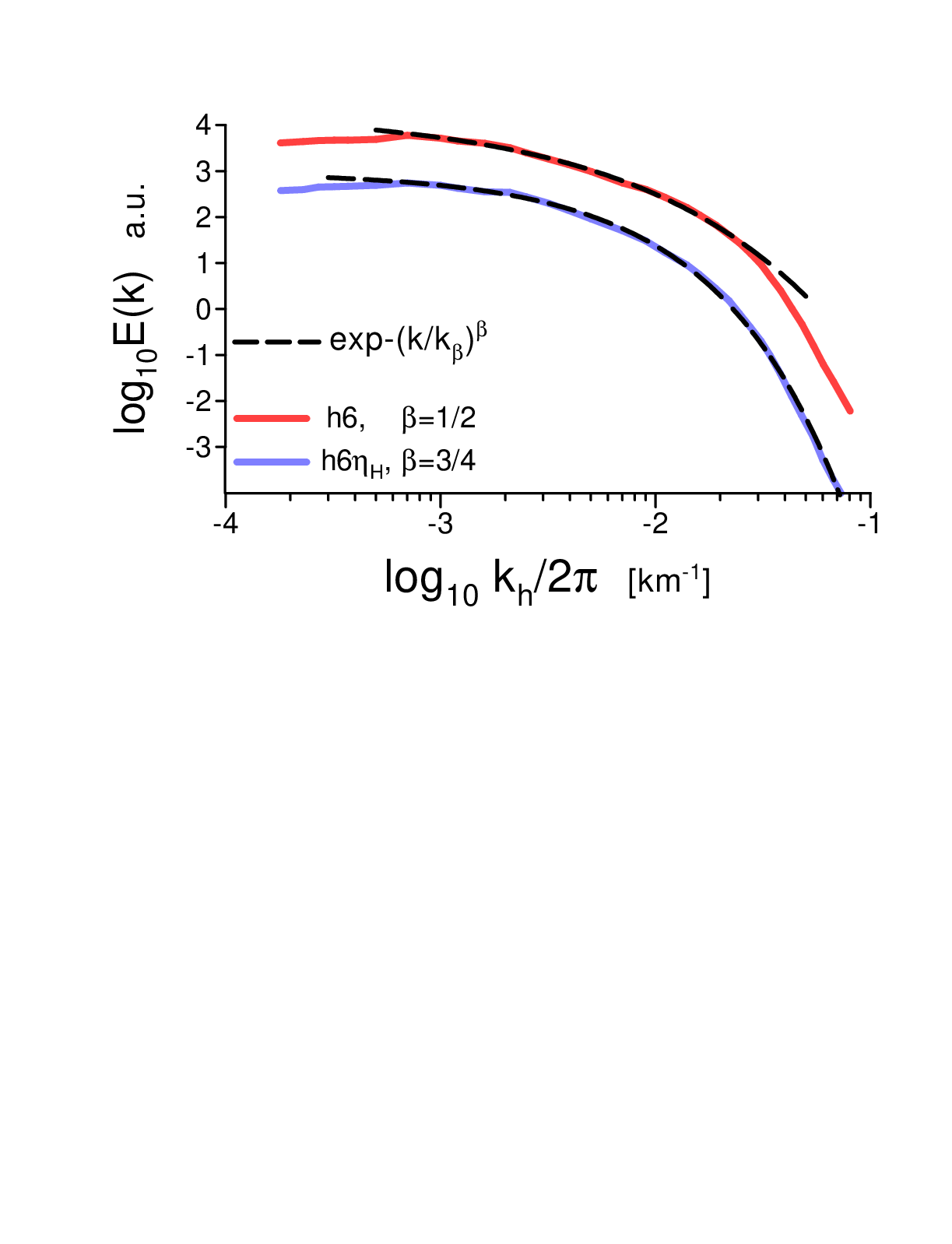} \vspace{-6.15cm}
\caption{Magnetic energy spectra averaged over a vertical range between $z = 0$ km and $z = -200$ km, and over time. The dissipative and non-dissipative cases correspond to the bottom and top curves respectively (DNS). The spectra are vertically shifted for clarity.} 
\end{figure}
 
\begin{figure} \vspace{-0.5cm}\centering \hspace{-0.8cm}
\epsfig{width=.48\textwidth,file=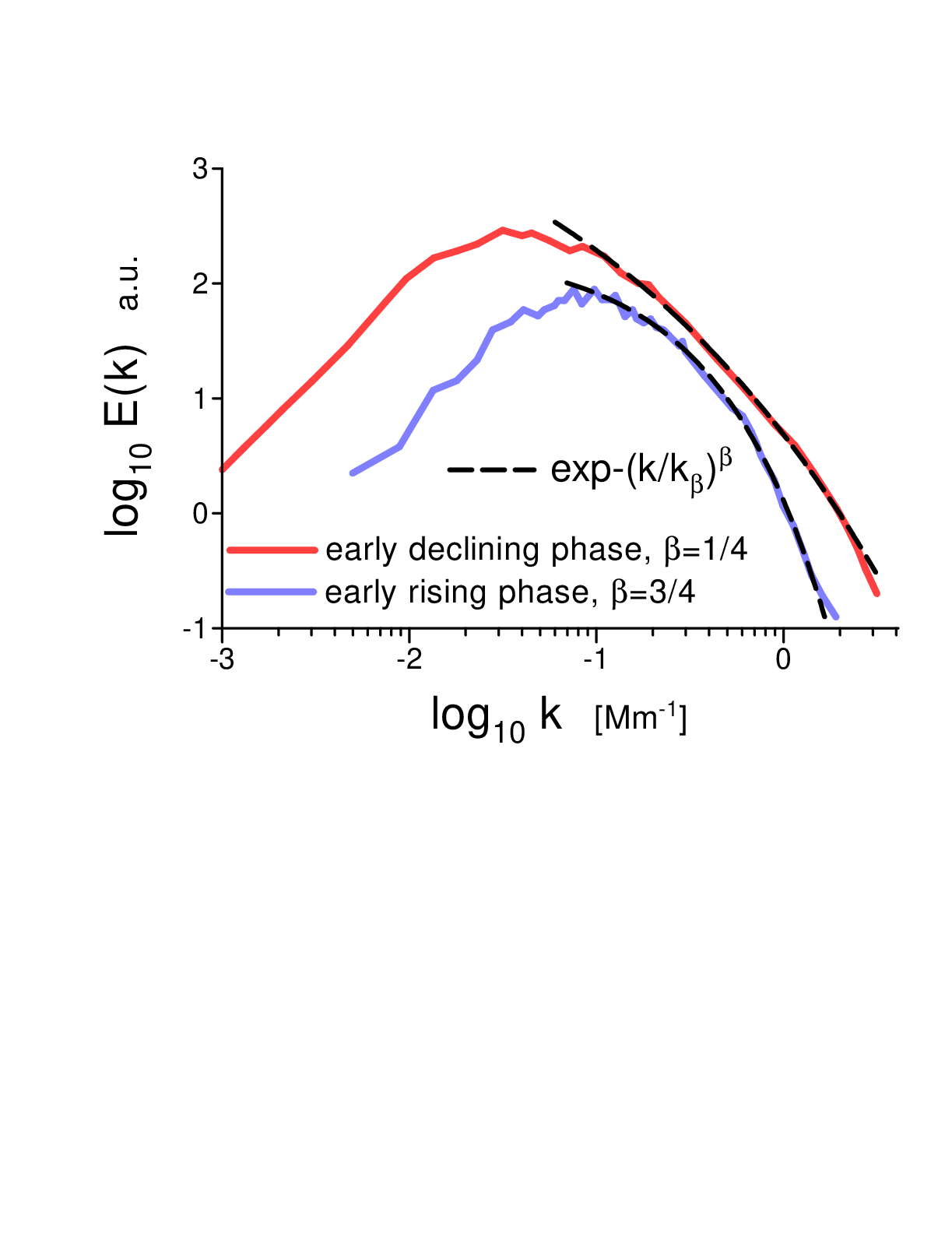} \vspace{-4.65cm}
\caption{Magnetic energy spectra for synoptic maps of the early rising (bottom) and early declining (top) phases of solar cycle 24. } 
\end{figure}

\begin{figure} \vspace{-0.5cm}\centering \hspace{-0.8cm}
\epsfig{width=.46\textwidth,file=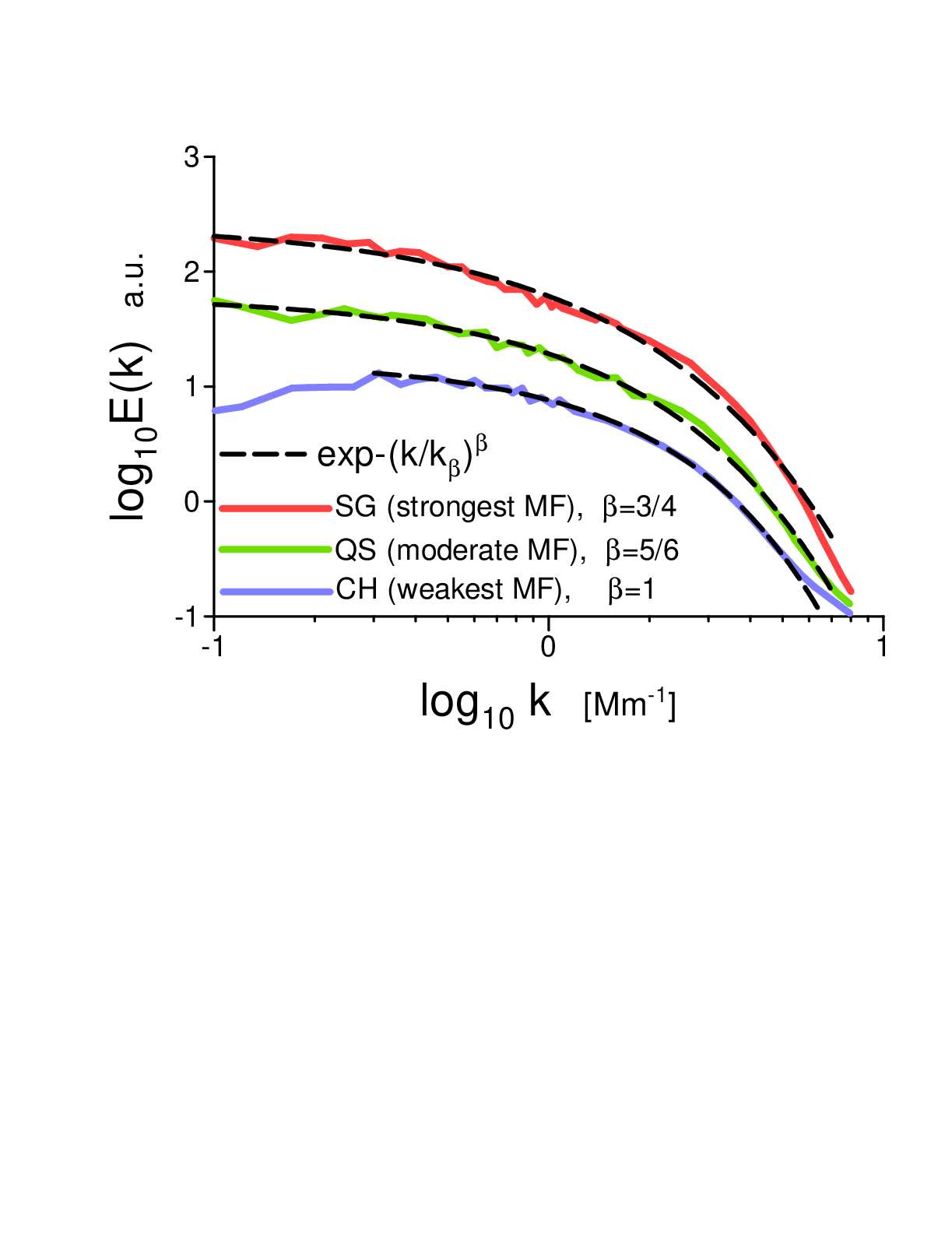} \vspace{-4.5cm}
\caption{Magnetic energy spectra in undisturbed photosphere: CH - coronal hole (bottom), QS - quiet Sun region, SG - supergranular network (top), MF - magnetic field. } 
\end{figure}

  Figure 18 shows the magnetic energy spectra for synoptic maps of the early rising and early declining phases of solar cycle 24. In the time of the early declining phase both magnetic energy and magnetic helicity have their maxima, whereas during the early rising phase, the magnetic energy has its minimum \cite{singh}. The spectral data were taken from Figs. 4.2b and 4.3b of Ref. \cite{prabhu}. The data used for computing the spectrum of the early declining phase were obtained by the Helioseismic and Magnetic Imager onboard Solar dynamics observatory, whereas the data used for computing the spectrum of the early rising phase were obtained by the Vector Spectromagnetograph on the ground-based Synoptic Optical Long-term Investigations of the Sun telescope. \\
  
  The dashed curves indicate the best fit of the spectra with Eq. (31) ($\beta =3/4$, dissipative distributed chaos dominated by the magnetic Birkhoff-Saffman invariant) for the early rising phase and Eq. (21) ($\beta =1/4$, the magneto-inertial range of scales) for the early declining phase when both magnetic energy and magnetic helicity have their maxima. \\ 
 
  Figure 19 shows magnetic energy spectra in an undisturbed (outside active regions) photospheric area computed using the results of measurements produced with the Helioseismic and Magnetic Imager onboard Solar Dynamic Observatory on June 19, 2017, in the center of the solar disk. The spectral data were taken from Fig. 4 of a paper Ref. \cite{ak}. The solar area consists of three subregions: a coronal hole (CH), a quiet Sun region (QS), and a supergranular network (SG). On the whole, it was an area of weak magnetic fields, but the CH, QS, and SG subregions had comparatively different intensities of their magnetic fields: CH - weakest, QS - moderate, and SG - strongest. The super-granular network (SG subregion) was an area of decayed several rotations ago NOAA 12242–12259 active regions. \\
    
    The dashed curves indicate the best fit of the spectra with the exponent Eq. (2) ($\beta =1$, deterministic chaos in the CH weak magnetic field), with the Eq. (33) ($\beta =5/6$, dissipative distributed chaos dominated by the magnetic Loitsianskii invariant in the QS moderate magnetic filed), and with the Eq. (31) ($\beta =3/4$, dissipative distributed chaos dominated by the magnetic Birkhoff-Saffman invariant in the SG strongest magnetic field). \\
    
    As it could be expected the value of the parameter $\beta$ decreases and, consequently the randomness of the magnetic field increases for the subregions with larger intensity of the magnetic field. \\
    
     Figure 2 shows the evolution of the magnetic energy spectra for a large emerging solar active region NOAA 11726. The spectral data were taken from Fig. 2a of the recent paper Ref. \cite{kka}. The measurements were provided by Helioseismic and Magnetic Imager located on board of the Solar Dynamics Observatory.  These spectra are well-fitted by the sequence of the spectral laws (the dashed curves): starting from the deterministic chaos Eq. (2) ($\beta =1$), then the dissipative distributed chaos Eq (31) ($\beta =3/4$), then the helical distributed chaos Eq. (9) ($\beta =1/2$), and then the magneto-inertial range of scales Eq. (21) ($\beta =1/4$, a precursor of the hard turbulence. \\
     
      Figure 20 shows the evolution of the magnetic energy spectra for a smaller emerging solar active region NOAA 11781. The spectral data were taken from Fig. 2b of the same paper Ref. \cite{kka}. The measurements were also provided by Helioseismic and Magnetic Imager located on board the Solar Dynamics Observatory. These spectra are well-fitted by the sequence of the analogous spectral laws (the dashed curves) and only the most energetic magneto-inertial range of scales Eq. (21) ($\beta =1/4$) is absent here (cf Fig. 2). \\
     
\begin{figure} \vspace{-0.9cm}\centering \hspace{-0.8cm}
\epsfig{width=.46\textwidth,file=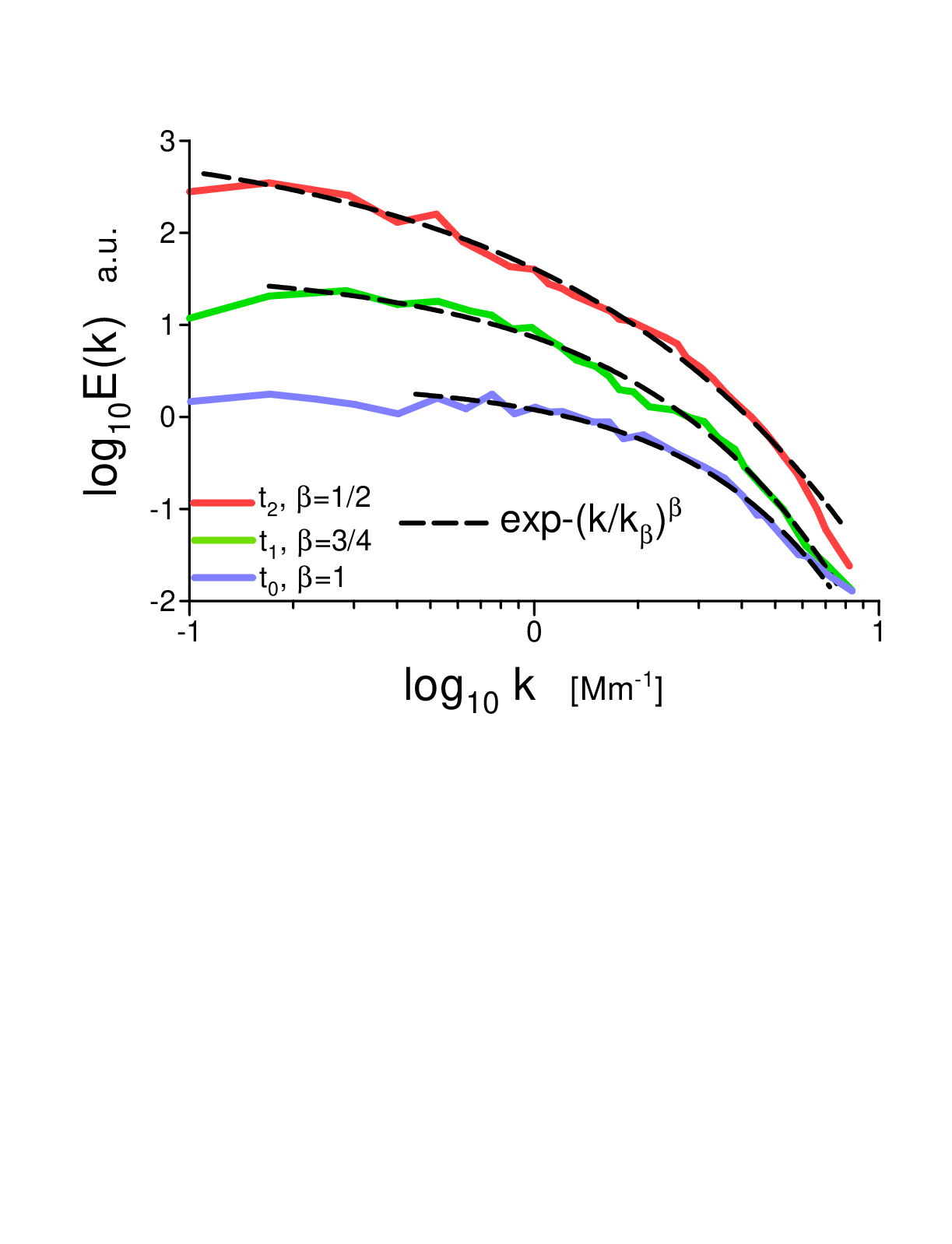} \vspace{-4.7cm}
\caption{Evolution of the magnetic energy spectra for an emerging solar active region NOAA 11781.  } 
\end{figure}
 
\begin{figure} \vspace{-0.45cm}\centering \hspace{-0.8cm}
\epsfig{width=.45\textwidth,file=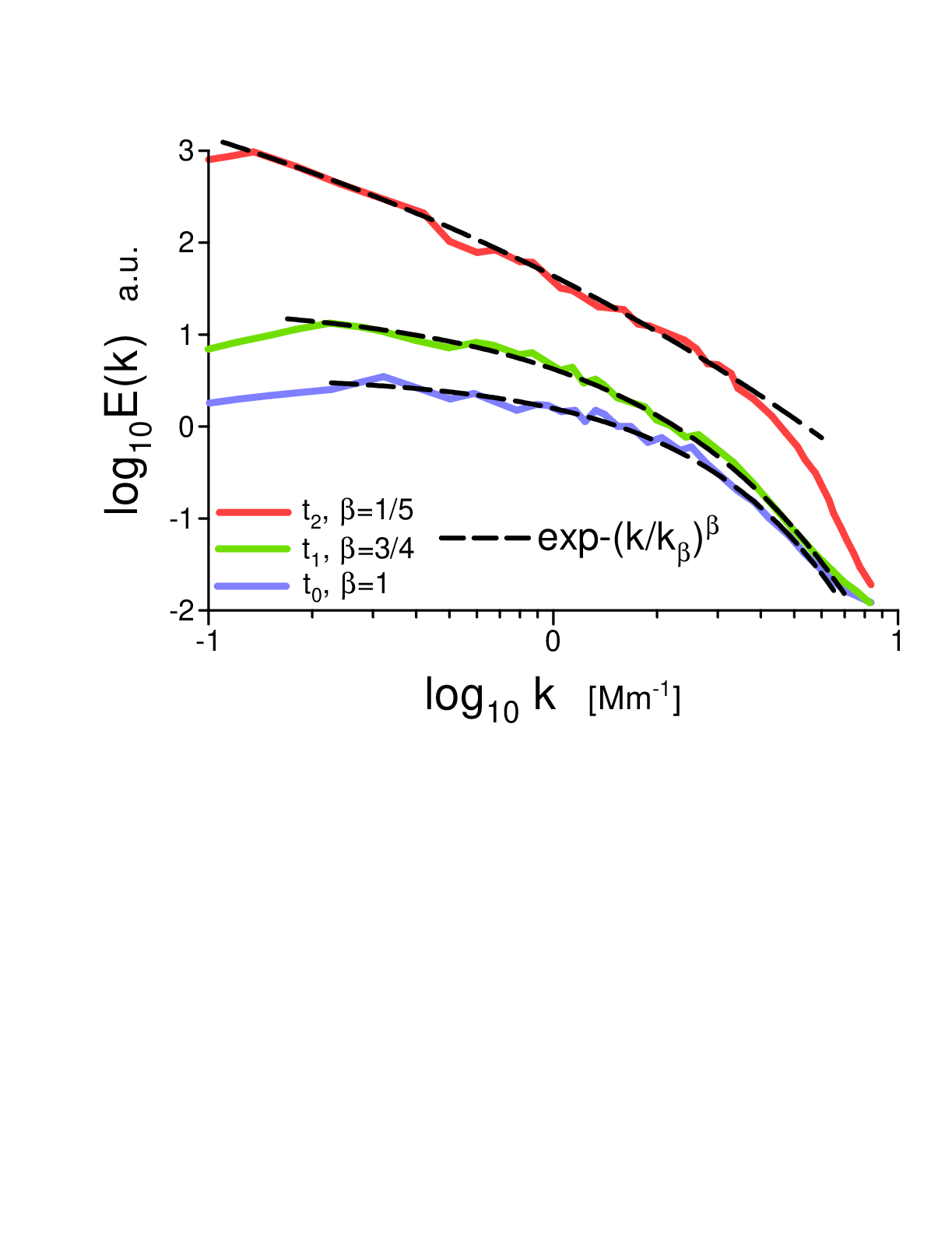} \vspace{-4.5cm}
\caption{Evolution of the magnetic energy spectra for an emerging solar active region NOAA 12219. } 
\end{figure}
 
\subsection{The role of a mean magnetic field}

  To take into account a mean magnetic field (if it is present) in the case of a magneto-inertial range of scales one should replace the energy dissipation rate $\varepsilon$ in the Eq. (16) by $(\varepsilon \widetilde{B}_0)$ (where  $\widetilde{B}_0 = B_0/\sqrt{\mu_0\rho}$ is normalized mean magnetic field - in the Alfvenic units, with the same dimension as velocity) \cite{ir}, and the $\varepsilon_h$ by the modified magnetic helicity (Eq. 4) dissipation rate $\varepsilon_{\hat h}$. \\
  
  The dimensional considerations result in
\begin{equation}
 B_c \propto \varepsilon_{\hat h}^{1/2}~ (\varepsilon \widetilde{B}_0)^{-1/8}  k_c^{1/8} 
 \end{equation} 
 i.e. the parameter $\alpha = 1/8$. Then from Eq. (20) we obtain $\beta =1/5$ 
\begin{equation}
 E(k) \propto \exp-(k/k_{\beta})^{1/5}  
 \end{equation}
 
    Figure 21 shows the evolution of the magnetic energy spectra for an emerging solar active region NOAA 12219. The spectral data were taken from Fig. 1 of a recent paper Ref. \cite{kak}. The solar magnetogram images used for the calculation of the spectra were obtained by the Helioseismic and Magnetic Imager onboard the Solar Dynamics Observatory spacecraft. The time $t_0$ corresponds to the onset of the active region emergence, $t_1$ corresponds to the time when only the first imprints of the emerging active region can be seen in the magnetogram, the time $t_2$ corresponds to the peak of magnetic flux. 
 
\begin{figure} \vspace{-1.1cm}\centering \hspace{-0.8cm}
\epsfig{width=.5\textwidth,file=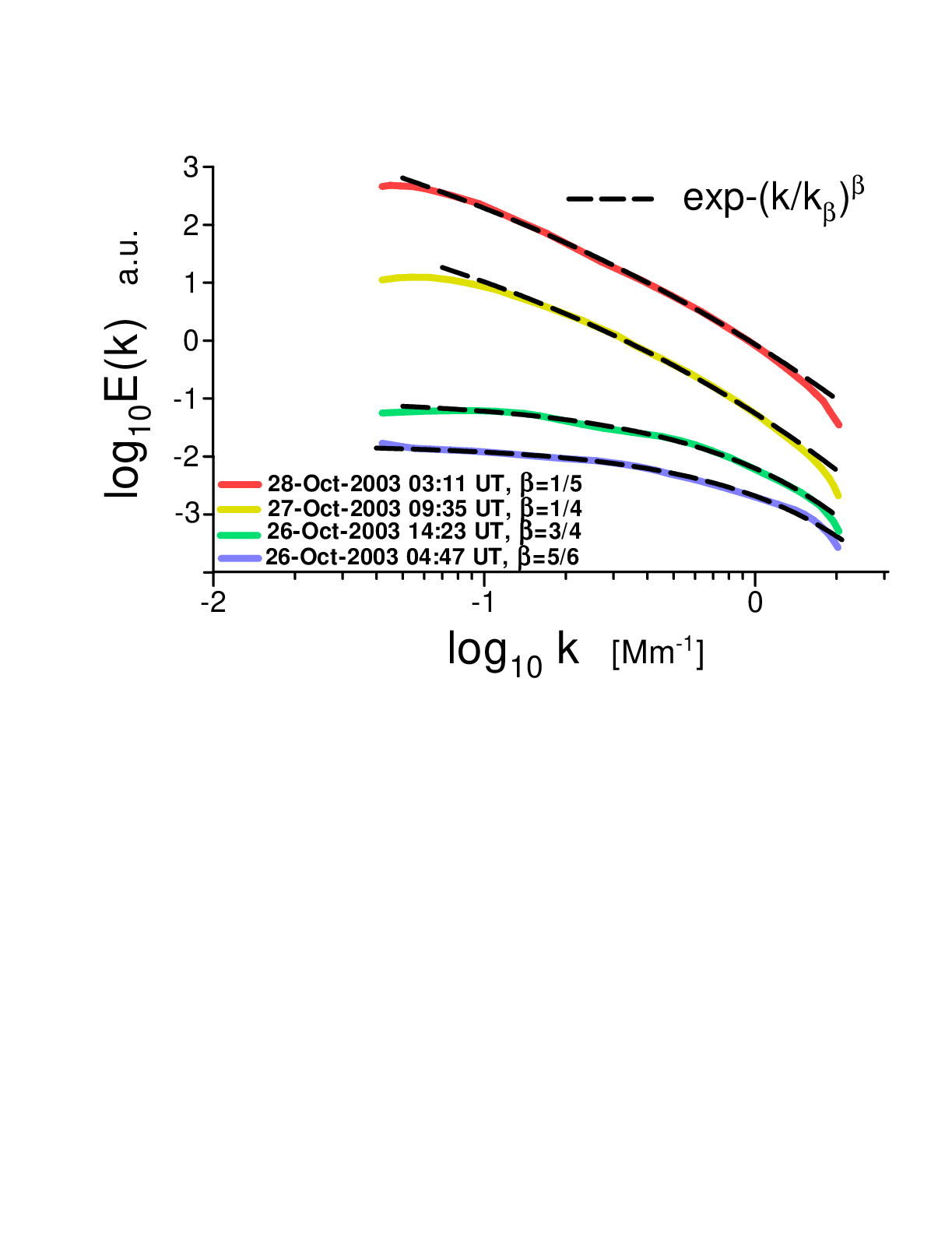} \vspace{-5.45cm}
\caption{Evolution of the magnetic energy spectra for a rapidly emerging solar active region AR NOAA 10488. The spectra are vertically shifted for clarity. } 
\end{figure}
    
    These spectra are fitted by the sequence of the spectral laws (the dashed curves): starting from the deterministic chaos Eq. (2) ($\beta =1$), then the dissipative distributed chaos Eq (31) ($\beta =3/4$), and then the magneto-inertial range of scales Eq. (43)  ($\beta =1/5$), corresponding to the magneto-inertial range under the strong influence of a mean magnetic field. \\
  
 Figure 22 shows the evolution of the magnetic energy spectra for a rapidly emerging active region NOAA 10488. The spectral data were taken from Fig. 8 of a paper Ref. \cite{hew}. The solar magnetogram images used for the calculation of the spectra were obtained by the Michelson Doppler Imager
onboard the Solar and Heliospheric Observatory. An M-class flare and two C-class flares occurred in the active region, and also an active region NOAA 10493 appeared near the AR  NOAA 10488 during the period of 27-28 October 2003.\\

  The dashed curves in Fig. 22 indicate the best fit of the spectra with the Eq. (33) ($\beta =5/6$) and Eq. (31) ($\beta =3/4$), corresponding to the dissipative distributed chaos dominated by the magnetic Loitsianskii and Birkhoff-Saffman invariants at the earlier stages of the AR NOAA 10488 emergence; and with the Eq. (21) ($\beta =1/4$) and Eq. (43) ($\beta =1/5$), corresponding to the magneto-inertial range of scales at the mature stage of the AR NOAA 10488 emergence (cf Fig. 2). 

 \section{Conclusions}
 
  While the chaotic/turbulent magnetohydrodynamic dynamo is a thriving subject of research in fluid dynamics (see for recent reviews Refs. \cite{md}, \cite{tob}, \cite{sheb} and references therein), theory of the process of randomization of generated by the small-scale dynamo magnetic field is still underdeveloped. The randomization process can be understood and quantified using the notion of distributed chaos. \\
  
  The given above numerous examples show that this process is controlled by the magnetohydrodynamic invariants for different types of flows. While at earlier stages of randomization, the dissipative invariants are the dominating ones for more advanced stages the magnetic helicity plays the dominant role directly or through the Komogorov-Iroshnikov types of phenomenology. The small-scale dynamo can coexist with the large-scale one.\\
  
  In the frames of this approach, the results of numerical simulations are in quantitative agreement with the geophysical and solar observations despite the considerable differences in the scales and physical parameters.

\section{Acknowledgments }

  I thank H.K. Moffatt, A. Pikovsky, and J.V. Shebalin for stimulating discussions.

\end{document}